\documentclass[prl,reprint,twocolumn,footinbib,longbibliography,superscriptaddress]{revtex4-2}
\usepackage{amsmath}
\usepackage{amsfonts}
\usepackage{amssymb}
\usepackage{lmodern}
\usepackage[dvipdfmx]{graphicx}
\usepackage[usenames,dvipsnames]{xcolor}
\usepackage{bm}
\usepackage{amsthm}
\usepackage[whole]{bxcjkjatype}

\usepackage{physics}
\usepackage{mathrsfs}

\usepackage{hyperref}

\newcommand{\xv}{[\mathbf{X}]}
\newcommand{\yv}{[\mathbf{Y}]}

\newcommand{\X}{\mathrm{X}}
\newcommand{\Y}{\mathrm{Y}}

\newcommand{\st}{\mathsf{S}}

\renewcommand{\L}{\mathcal{L}}
\newcommand{\C}{\mathcal{C}}
\newcommand{\A}{\mathrm{A}}

\newcommand{\mr}[1]{\mathrm{#1}}
\renewcommand{\S}{\mathscr{S}}

\begin{document}

\title{Thermodynamic uncertainty relation and thermodynamic speed limit in deterministic chemical reaction networks}
\date{\today}

\author{Kohei Yoshimura}
\affiliation{Department of Physics, The University of Tokyo, 7-3-1 Hongo, Bunkyo-ku, Tokyo 113-0031, Japan}
\author{Sosuke Ito}
\affiliation{Department of Physics, The University of Tokyo, 7-3-1 Hongo, Bunkyo-ku, Tokyo 113-0031, Japan}
\affiliation{JST, PRESTO, 4-1-8 Honcho, Kawaguchi, Saitama, 332-0012, Japan}

\begin{abstract}
    We generalize thermodynamic uncertainty relation (TUR) and thermodynamic speed limit (TSL) for deterministic chemical reaction networks (CRNs). 
    The scaled diffusion coefficient derived by considering the connection between macroscopic CRNs and mesoscopic CRNs plays an essential role in our results. 
    The TUR shows that the product of the entropy production rate and the ratio of the scaled diffusion coefficient to the square of the rate of concentration change is bounded below by 2. 
    The TSL states a trade-off relation between speed and thermodynamic quantities, the entropy production and the time-averaged scaled diffusion coefficient. 
    The results are proved under the general setting of open and non-ideal CRNs. 
\end{abstract}

\maketitle

\textit{Introduction.---}
\begin{figure*}
    \centering
    \includegraphics[width=\hsize]{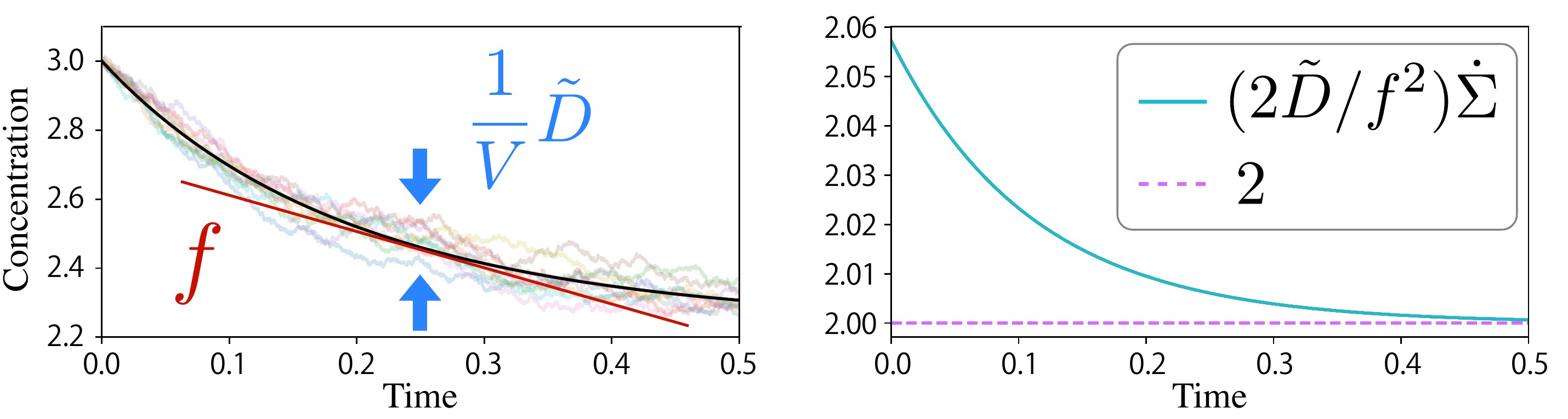}
    \caption{Schematic diagram of thermodynamic uncertainty relation in chemical thermodynamics [Eq.~\eqref{TUR}]. 
    The left graph shows the concentrations of $\A$ in a reaction $2\A\rightleftharpoons\mathrm{B}$ calculated by the chemical Langevin equation corresponding to the Fokker--Planck equation~\eqref{fpeq}. 
    The dark curve is obtained with the volume $V$ set $10^7$ times as great as that for the light curves. 
    Although the same number of curves are plotted for each $V$, the curves concentrate on the single curve when $V$ is large. 
    By taking the thermodynamic limit in such a way, we have a rigorous inequality between the changing rate of the concentration $f$, the scaled diffusion coefficient $\tilde{D}$, and the entropy production rate $\dot{\Sigma}$, namely, the thermodynamic uncertainty relation, as shown in the right figure.}
    \label{fig_concept}
\end{figure*}
It has been a fundamental question whether there are universal laws in nonequilibrium systems or processes like equilibrium thermodynamics. 
In the last two decades, our understanding of the thermodynamic structure of nonequilibrium mesoscopic systems has been substantially gained with the aid of stochastic thermodynamics~\cite{seifert2012stochastic,sekimoto2010stochastic}. 

The following are two examples of discoveries made by stochastic thermodynamics.
One is thermodynamic uncertainty relation (TUR)~\cite{barato2015thermodynamic,horowitz2020thermodynamic}.
A TUR states a trade-off relation between a relative fluctuation and dissipation. The former is typically evaluated by the diffusion constant $D$ and a current $J$ as $2D/J^2$, while the latter is given by the entropy production rate $\sigma$. The original TUR shows the trade-off relation by an inequality $\mathcal{Q}:=(2D/J^2)\sigma\geq 2$~\cite{barato2015thermodynamic}. 
Subsequently, various variants in mesoscopic systems~\cite{gingrich2016dissipation,pietzonka2016universal,horowitz2017proof,proesmans2017discrete,pietzonka2017finite, dechant2018current,dechant2018multidimensional,pietzonka2018universal, timpanaro2019thermodynamic,hasegawa2019fluctuation,falasco2020unifying,wolpert2020uncertainty,otsubo2020estimating,manikandan2020inferring,liu2020thermodynamic} 
and an extension to quantum systems~\cite{hasegawa2021thermodynamic} have been developed.
The other example is thermodynamic speed limit (TSL)~\cite{shiraishi2018speed,funo2019speed,van2020unified,falasco2020dissipation}.
A TSL gives a lower bound to the time it takes for a system to change using thermodynamic quantities such as the entropy production. 
TSLs typically indicate a universal trade-off between speed and dissipation. 
Speed limits (SLs) were originally developed for microscopic systems described by quantum mechanics~\cite{mandelstam1945uncertainty,margolus1998maximum}. To date, many SLs have been found by using mathematically elaborated methods~\cite{aurell2012refined,pires2016generalized,okuyama2018quantum,ito2018stochastic,dechant2019thermodynamic,ito2020stochastic,gupta2020tighter,nicholson2020time,yoshimura2021information,van2021geometrical,nakazato2021geometrical}. 

However, the universal thermodynamic principles, TUR and TSL, have been restricted to mesoscopic or microscopic systems described by stochastic thermodynamics or quantum mechanics. 
Whether such principles hold in other nonequilibrium systems like deterministic chemical reaction networks (CRNs) is nontrivial. 
Chemical thermodynamics has been an essential thermodynamic theory of nonequilibrium systems before the birth of stochastic thermodynamics~\cite{gibbs1878equilibrium,de1936thermodynamic,kondepudi2014modern}. 
Remarkably, the original derivation of TUR is obtained in a stochastic model of enzymatic reaction~\cite{barato2015thermodynamic}. 
Mesoscopic theory of chemical reactions can be described by stochastic thermodynamics~\cite{ge2016mesoscopic,ge2017mathematical}, while macroscopic theory, e.g., thermodynamic theory of biochemical reaction networks, is not~\cite{kondepudi2014modern,beard2008chemical}. 
Because of their deep connection~\cite{kurtz1972relationship,kurtz1978strong}, wisdom of stochastic thermodynamics is still useful for macroscopic chemical thermodynamics~\cite{ge2016nonequilibrium,rao2016nonequilibrium,falasco2018information,wachtel2018thermodynamically,avanzini2020thermodynamics,yoshimura2021information,avanzini2021nonequilibrium}. 
However, the latest knowledge of stochastic thermodynamics such as TUR and TSL has not been sufficiently considered in chemical thermodynamics.

In this letter, we obtain a TUR and a TSL in deterministic CRNs by focusing on a relationship between the mesoscopic and macroscopic theory of chemical reactions. 
In both TUR and TSL, the intrinsic fluctuations of CRNs play an important role. The fluctuations in chemical reactions get smaller when the size of the system increases as shown in Fig.~\ref{fig_concept}. However, they can be considered in macroscopic CRNs by scaling by the volume (the scaled diffusion coefficient $\tilde{D}$ in Fig.~\ref{fig_concept}). 
We obtain a TUR between this measure of fluctuations $\tilde{D}$, the rate of concentration change, and the entropy production rate. 
This measure of fluctuations is also important in the TSL, which shows a relation between speed and thermodynamic quantities, the scaled diffusion coefficient and the entropy production. 
These results are proved under highly general settings used in recent studies~\cite{rao2016nonequilibrium, avanzini2021nonequilibrium}. 
We illustrate the TUR and TSL in concrete models of CRNs. 

\textit{Setup.---}
We examine open CRNs that consist of $N+N'$ chemical species. Within them, we assume that the concentrations of $N'$ species are controlled externally. We denote the $N$ kinds of internal species by $\X_i\;(i\in \mathscr{S}_\X:=\{1,\dots,N\})$, and denote the other chemostatted species by $\Y_i\;(i\in \mathscr{S}_\Y:=\{N+1,\dots,N+N'\})$. 
Here, we define $\mathscr{S}_\X$ and $\mathscr{S}_\Y$ as the index sets of the two kinds of species. 
We may use $\alpha$ to collectively represent $\X$ and $\Y$. That is, $\alpha_i$ means $\X_i$ if $i\in \mathscr{S}_\X$, and vice versa. 
Chemical reaction networks have $M$ reversible reactions labelled by $\rho\in \mathscr{R}:=\{1,\dots,M\}$. Each reaction has two directions of reaction since it is reversible. We call one of the two the forward reaction and the other the backward reaction. We denote the number of $\alpha_i$ involved in the $\rho$th forward reaction by $\nu_{i\rho}^\alpha$, and that involved in the backward reaction by $\kappa^\alpha_{i\rho}$. Then, the $\rho$th reaction can be written as follows:
\begin{align}
    \sum_{i\in \mathscr{S}_\X} \nu^\mathrm{X}_{i\rho}\mathrm{X}_i
    +\sum_{i\in \mathscr{S}_\Y} \nu^\mathrm{Y}_{i\rho}\mathrm{Y}_i
    \rightleftharpoons
    \sum_{i\in \mathscr{S}_\X} \kappa^\mathrm{X}_{i\rho}\mathrm{X}_i
    +\sum_{i\in \mathscr{S}_\Y} \kappa^\mathrm{Y}_{i\rho}\mathrm{Y}_i.
\end{align}
For both the internal species and the chemostatted species, we define the respective stoichiometric coefficient matrix $\st^\alpha$ by $\st_{i\rho}^\alpha:=\kappa_{i\rho}^\alpha-\nu_{i\rho}^\alpha$. 
Each element $\st_{i\rho}^\alpha$ gives the net increase (resp. decrease) in $\alpha_i$ molecule in the $\rho$th forward (resp. backward) reaction. Combining them, we can obtain the total stoichiometric coefficient matrix $\st=((\st^\X)^\mathrm{T}\;(\st^\Y)^\mathrm{T})^\mathrm{T}$, where the superscript $\mathrm{T}$ represents the transposition. 

We also consider the kinetics of CRNs. 
Let $[\bm{\alpha}]_t=([\alpha_i]_t)_{i\in\mathscr{S}_{\alpha}}$ denote the concentrations of $\alpha_i$'s at time $t$. 
Throughout this paper, we only consider homogeneous CRNs where the concentrations do not depend on the position. 
Let the rate of the $\rho$th reaction be $J_\rho=J_\rho^+-J_\rho^-$, where $J_\rho^+$ (resp. $J_\rho^-$) is the reaction rate of the forward (resp. backward) reaction. They are functions of the concentrations. 
Then, the kinetics of the concentrations are given by the rate equation: 
\begin{align}
    \dv{\xv_t}{t}=\st^\X\bm{J},\quad
    \dv{\yv_t}{t}=\st^\Y\bm{J}+\bm{\mathcal{J}}^\Y \label{rateeq_maintext}
\end{align}
where $\bm{J}=(J_\rho)_{\rho\in\mathscr{R}}$ is the vector of reaction rates and $\bm{\mathcal{J}}^\Y=(\mathcal{J}_i^\Y)_{i\in \mathscr{S}_\Y}$ is the vector of external flows  to control the concentrations of the chemostatted species.

We introduce thermodynamic structure to CRNs. 
To this end, we adopt the local detailed balance condition introduced in Ref.~\cite{avanzini2021nonequilibrium}: 
\begin{align}
    -(\bm{\mu}^\mathrm{T}\st)_\rho=RT\ln\frac{J_\rho^+}{J_\rho^-}, \label{wdb}
\end{align}
where $\bm{\mu}=(\mu_i)_{i\in \mathscr{S}_\X\cup \mathscr{S}_\Y}$ is the chemical potential, $(\cdot)_\rho$ is the $\rho$th element of the vector, $R$ is the gas constant, and $T$ is the temperature.
This is a core assumption when extending the framework of chemical thermodynamics to non-ideal systems~\cite{avanzini2021nonequilibrium}.
Because of the local detailed balance condition, the entropy production rate of chemical reactions is given as follows~\cite{schnakenberg1976network,rao2016nonequilibrium,avanzini2021nonequilibrium}:
\begin{align}
    \dot{\Sigma}=R\sum_{\rho\in\mathscr{R}} J_\rho\ln\frac{J_\rho^+}{J_\rho^-}\geq 0, \label{epr}
\end{align}
where the inequality is obtained since the signs of $J_\rho=J_\rho^+-J_\rho^-$ and $\ln(J_\rho^+/J_\rho^-)$ are the same, and it expresses the second law of thermodynamics. 
The total entropy production during a time interval $[0,\tau]$ is given by integrating the entropy production rate as $\Sigma :=\int_0^\tau \dd{t} \dot{\Sigma}$. 
Hereafter, we set $R=1$. 

In addition to the entropy production that involves all the reactions, we formally introduce partial entropy productions for specific chemical species. 
To define partial entropy productions, we define a subset of $\mathscr{R}$ for each subset of species $\mathscr{S}\subset\mathscr{S}_\X\cup\mathscr{S}_\Y$ by 
$\mathscr{R}_\mathscr{S}:=\{\rho\in\mathscr{R}\mid\exists i\in\mathscr{S},\;\mathsf{S}_{i\rho}\neq 0\}$. 
Next, we define the partial entropy production rate for a subset of chemical species $\mathscr{S}$ by
\begin{align}
    \dot{\Sigma}_\mathscr{S}:=\sum_{\rho\in\mathscr{R}_\mathscr{S}}J_\rho\ln\frac{J_\rho^+}{J_\rho^-}.
\end{align}
The partial entropy production is given by integrating the partial entropy production rate $\Sigma_\mathscr{S}:=\int_0^\tau\dd{t}\dot{\Sigma}_\mathscr{S}$. 
If $\mathscr{S}$ is a subset of $\mathscr{S}'$, $\mathscr{R}_\mathscr{S}$ is also a subset of $\mathscr{R}_{\mathscr{S}'}$. Thus, if $\mathscr{S}\subset\mathscr{S}'$, $\dot{\Sigma}_\mathscr{S}\leq \dot{\Sigma}_{\mathscr{S}'}$ and $\Sigma_\mathscr{S}\leq \Sigma_{\mathscr{S}'}$ hold. 
When $\mathscr{S}$ has only one element $\alpha_i$, we may substitute $i$ for $\mathscr{S}$ like $\dot{\Sigma}_i$ or $\mathscr{R}_i$. 

\textit{Main results.---}
We first state and prove the most important inequality for the derivation of our results: 
\begin{align}
    |f_i|
    \leq \sqrt{\tilde{D}_{ii}\dot{\Sigma}_i}, \label{mainineq}
\end{align}
where $f_i:=\sum_{\rho\in\mathscr{R}}\st_{i\rho}J_\rho$ and $\tilde{D}_{ii}:=(1/2)\sum_{\rho\in\mathscr{R}}\st_{i\rho}^2(J_\rho^++J_\rho^-)$. 
We note that the range of summation in the definition of $f_i$ and $\tilde{D}_{ii}$ can be replaced by the subset $\mathscr{R}_i$, namely, $f_i=\sum_{\rho\in\mathscr{R}_i}\st_{i\rho}J_\rho$ and $\tilde{D}_{ii}=(1/2)\sum_{\rho\in\mathscr{R}_i}\st_{i\rho}^2(J_\rho^++J_\rho^-)$, because $\st_{i\rho}=0$ if $\rho\notin\mathscr{R}_i$. 
This inequality is shown as follows. 
From the Cauchy--Schwarz inequality, we find
\begin{align}
    \Bigg|\sum_{\rho\in\mathscr{R}_i}\st_{i\rho}J_\rho\Bigg|
    &=\Bigg|\sum_{\rho\in\mathscr{R}_i}\st_{i\rho}\sqrt{J_\rho^++J_\rho^-}\frac{J_\rho}{\sqrt{J_\rho^++J_\rho^-}}\Bigg|\\
    &\leq \sqrt{\sum_{\rho\in\mathscr{R}_i} \st_{i\rho}^2(J_\rho^++J_\rho^-)}
    \sqrt{\sum_{\rho\in\mathscr{R}_i}\frac{J_\rho^2}{J_\rho^++J_\rho^-}}. \label{ineq9}
\end{align}
By using an inequality $2(a-b)^2/(a+b)\leq(a-b)\ln(a/b)$ that holds for any nonnegative real numbers $a,b$, we have
\begin{align}
    \sum_{\rho\in\mathscr{R}_i}\frac{J_\rho^2}{J_\rho^++J_\rho^-}
    \leq \frac{1}{2}\sum_{\rho\in\mathscr{R}_i}J_\rho\ln\frac{J_\rho^+}{J_\rho^-} =\frac{1}{2}\dot{\Sigma}_i. 
\end{align}
By combining these inequalities, we can obtain Eq.~\eqref{mainineq}. 

We next show how this inequality readily leads to a TUR. 
Our results are completely described by macroscopic quantities such as reaction rates, but the quantities appearing in Eq.~\eqref{mainineq}, $f_i$ and $\tilde{D}_{ii}$, should be understood from the mesoscopic point of view. 
Here, we assume $\dv*{[\mathbf{Y}]_t}{t}=\bm{0}$, but this assumption does not lose the generality of our discussion. 
When the stochasticity of reactions is strong, chemical reactions are described as Markov jump processes~\cite{gillespie1992rigorous}. By taking the thermodynamic limit, we can remove all the effects of the noise to recover the rate equation~\cite{kurtz1972relationship,kurtz1978strong}. If we leave the lowest-order noise, we have the chemical Fokker--Planck equation~\cite{gillespie2000chemical}: 
\begin{equation}
    \begin{split}
        &\pdv{p(t,\bm{x})}{t}\\&=-\sum_{i\in\mathscr{S}_\X}\pdv{x_i}\bqty{f_i p(t,\bm{x})}+\frac{1}{V}\sum_{i,j\in\mathscr{S}_\X}\pdv[2]{}{x_i}{x_j}[\tilde{D}_{ij}p(t,\bm{x})],
    \end{split} \label{fpeq}
\end{equation}
where $\bm{x}$ is the random variable that corresponds to the concentration, and $V$ is the volume as an expanding parameter.
This chemical Fokker--Planck equation has $f_i$ as the deterministic drift, and $V^{-1}\tilde{D}_{ij}=(2V)^{-1}\sum_{\rho\in\mathscr{R}}\st_{i\rho}\st_{j\rho}(J_\rho^++J_\rho^-)$ as the diffusion coefficient matrix (for derivation, see Supplemental Material~\cite{supplement}). 
Thus, $f_i$ and $\tilde{D}_{ii}$ can be seen as the measures of drifts and fluctuations that the CRN intrinsically has.
We call this scaled diffusion coefficient $\tilde{D}_{ij}$ simply the diffusion coefficient. 
As well as the ratio of the diffusion constant to the square of a current $2D/J^2$ that appears in the conventional TUR, the ratio of the diffusion coefficient to the square of the drift ${2\tilde{D}_{ii}}/{f_i^2}$ represents a relative fluctuation of the CRN. 
Therefore, the following relation can be seen as a thermodynamic uncertainty relation in chemical reactions: 
\begin{align}
    \pqty{\min_{i\in \mathscr{S}_\X\cup \mathscr{S}_\Y}\frac{2\tilde{D}_{ii}}{f_i^2}}\dot{\Sigma}\geq 2, \label{TUR}
\end{align}
where chemostatted species are reintroduced because the inequality in Eq.~\eqref{mainineq} holds for all $i\in \mathscr{S}_\X\cup \mathscr{S}_\Y$. 
This inequality is our first result. 
It is obtained from Eq.~\eqref{mainineq} and the fact that $\dot{\Sigma}_i\leq \dot{\Sigma}$.
It shows the trade-off relation between the entropy production rate $\dot{\Sigma}$ and the minimum of the relative fluctuation of chemical reactions $\min_{i\in \mathscr{S}_\X\cup \mathscr{S}_\Y}2\tilde{D}_{ii}/f_i^2$. 
As long as the local detailed balance condition~\eqref{wdb} is satisfied, it holds in any homogeneous CRNs, even if they are open, non-ideal, and non-stationary; thus, it is a universal law of chemical reactions. 


By integrating the inequality in Eq.~\eqref{mainineq}, we obtain a TSL of CRNs similar to the ones that have already been known in stochastic thermodynamics~\cite{shiraishi2018speed,van2020unified}. 
The following inequality is our second main reslut: 
\begin{align}
    \tau\geq \frac{L_\mathscr{S}(\xv_0,\xv_\tau)^2}{\langle\tilde{D}_\mathscr{S}\rangle_\tau\Sigma_\mathscr{S}}=:\tau_\mathscr{S}, \label{speedlimit}
\end{align}
where $\mathscr{S}$ is a subset of $\mathscr{S}_\X$ that has $|\mathscr{S}|$ elements,  $L_\mathscr{S}(\xv_t,\xv_{t'}):=|\mathscr{S}|^{-1}\sum_{i\in\mathscr{S}}|[\X_i]_t-[\X_i]_{t'}|$, $\tilde{D}_\mathscr{S}$ is the average of the diagonal elements of the diffusion coefficient matrix with respect to $\mathscr{S}$ given by $\tilde{D}_\mathscr{S}:=|\mathscr{S}|^{-1}\sum_{i\in\mathscr{S}} \tilde{D}_{ii}$, and the bracket represents the time average $\langle\tilde{D}_\mathscr{S}\rangle_\tau:=\tau^{-1}\int_0^\tau \dd{t}\tilde{D}_\mathscr{S}$. 
This inequality indicates a trade-off relation between speed and other physical quantities, the diffusion coefficient and the entropy production. 
It gives a lower bound to the time needed for a concentration distribution to change into another one. 
It shows that the time average of the diffusion coefficient or the entropy production must be increased when one tries making the time shorter by controlling external parameters. 
In particular, if the diffusion coefficient does not depend on parameters so much, the entropy production will be the complementary quantity to the changing speed. We will demonstrate this trade-off relation by a numerical calculation.

We prove the TSL. 
Because of the rate equation $\dv*{[\X_i]_t}{t}=f_i$, we have $\abs{[\X_i]_0-[\X_i]_\tau}=\abs{\int_0^\tau\dd{t}f_i}$. From the triangle inequality and the inequality in Eq.~\eqref{mainineq}, we find 
\begin{align}
    \abs{\int_0^\tau\dd{t}f_i}
    \leq \int_0^\tau\dd{t}\abs{f_i}
    \leq\abs{\int_0^\tau\dd{t}\sqrt{\tilde{D}_{ii}\dot{\Sigma}_i}}. 
\end{align}
By using the Cauchy--Schwarz inequality, we see that it is bounded as
\begin{align}
    \abs{\int_0^\tau\dd{t}\sqrt{\tilde{D}_{ii}\dot{\Sigma}_i}}
    \leq 
    \sqrt{\int_0^\tau\dd{t}\tilde{D}_{ii}}
    \sqrt{\int_0^\tau\dd{t}\dot{\Sigma}_i}
    =\sqrt{\tau\langle\tilde{D}_{ii}\rangle_\tau\Sigma_i}. 
\end{align}
Taking summation for $i\in\mathscr{S}$ leads to 
\begin{align}
    \sum_{i\in\mathscr{S}}\abs{[\X_i]_0-[\X_i]_\tau}
    &\leq\sqrt{\tau}\sum_{i\in\mathscr{S}}\sqrt{\langle\tilde{D}_{ii}\rangle_\tau\Sigma_i}\\
    &\leq \sqrt{\tau\Sigma_\mathscr{S}}\sum_{i\in\mathscr{S}}\sqrt{\langle\tilde{D}_{ii}\rangle_\tau},
\end{align}
where we use the fact that $\{i\}\subset\mathscr{S}$, so $\Sigma_i\leq \Sigma_\mathscr{S}$. 
The Cauchy--Schwarz inequality finally yields the following inequality: 
\begin{align}
    \sum_{i\in\mathscr{S}}\abs{[\X_i]_0-[\X_i]_\tau}
    \leq \sqrt{\tau\Sigma_\mathscr{S}}\sqrt{N_\mathscr{S}\sum_{i\in\mathscr{S}}\langle\tilde{D}_{ii}\rangle_\tau}. 
\end{align}
This inequality is readily turned into the form of Eq.~\eqref{speedlimit}. 

\textit{Example of the TUR.---}
\begin{figure}
    \centering
    \includegraphics[width=\hsize]{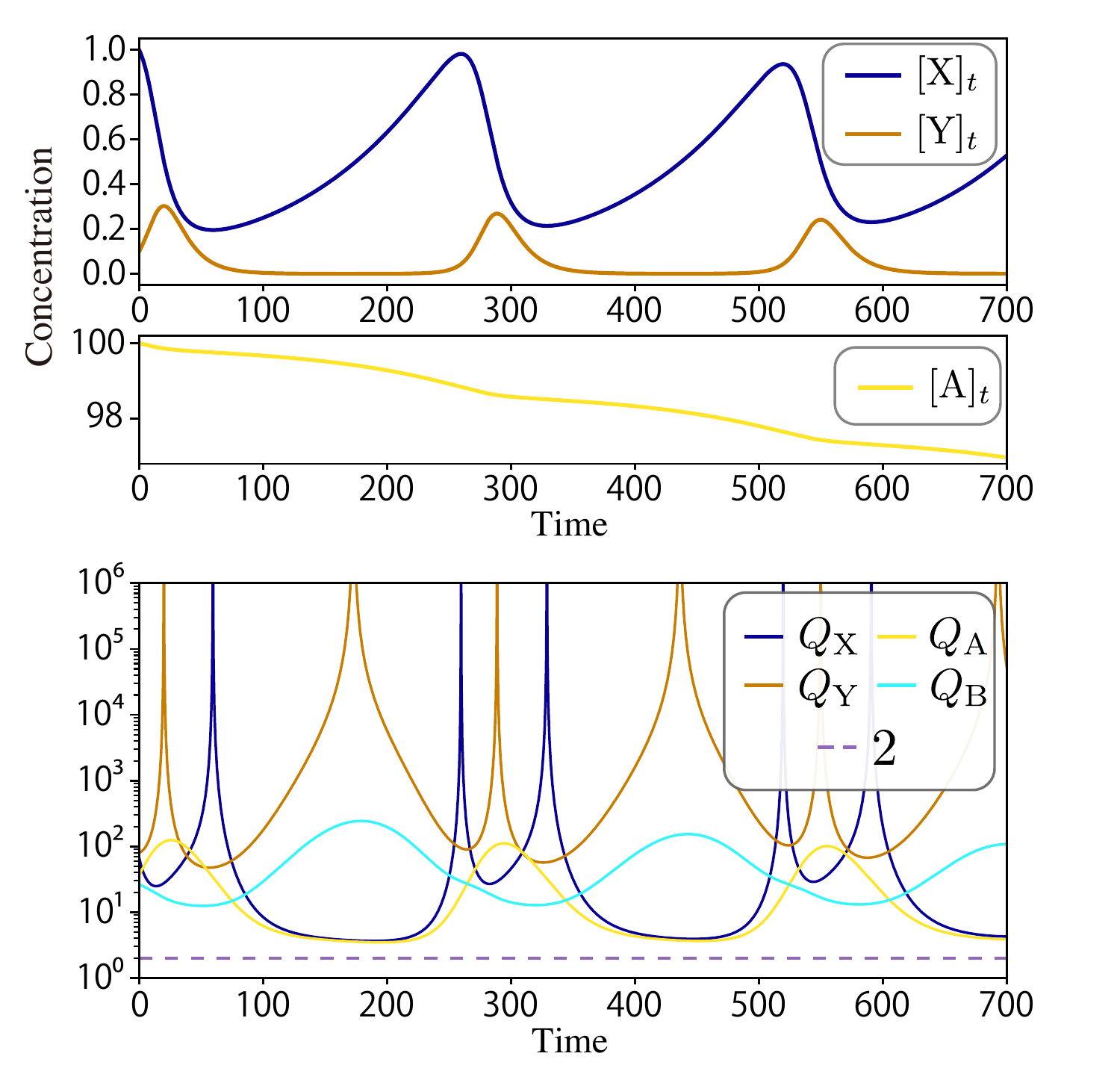}
    \caption{Concentration changes of $\X$, $\Y$, and $\A$ in the damped Lotka--Volterra model [Eq.~\eqref{LotkaVolterra}] are shown in the upper panels. The former two oscillate, while that of $\A$ monotonically decreases.
    Confirmation of the TUR~\eqref{TUR} is done in the lower panel. 
    $Q_i$'s are always bounded below by $2$. 
    Those of oscillating species, $\X,\Y$, are larger than that of the monotonically changing species $\A$ on average. }
    \label{fig_LV}
\end{figure}
We illustrate the TUR through a model of open oscillatory CRN. We consider the following damped Lotka--Volterra chemical reaction model~\cite{strogatz2018nonlinear}: 
\begin{align}
    \begin{aligned}
    \X+\mathrm{A}\rightleftharpoons 2\X,\quad
    \X+\Y\rightleftharpoons 2\Y,\quad
    \Y\rightleftharpoons\mathrm{B},
    \end{aligned}\label{LotkaVolterra}
\end{align}
where we set the concentration of $\mathrm{B}$ constant, so it is a model of open CRN. 
We numerically solve the rate equation, assuming that the reaction rates are given by the mass-action law (for details, see Supplemental Material~\cite{supplement}). 
As shown in the upper two panels in Fig.~\ref{fig_LV}, the concentrations of $\X$ and $\Y$ oscillate while that of $\mathrm{A}$ monotonically decreases. 

In the lower panel of Fig.~\ref{fig_LV}, we exhibit $Q_i:=(2\tilde{D}_{ii}/f_i^2)\dot{\Sigma}$ for $i\in\{\mathrm{X,Y,A,B}\}$. 
They are bounded below by $2$ shown by the purple dashed line. 
On average, $Q_i$'s of the oscillating species are bigger than $Q_\A$. 
That is because $f_i$'s of oscillating species oscillate around zero and are smaller on average than $f_\A$, while $\tilde{D}_{ii}$'s take nonzero values that are the same order as the diffusion coefficient of $\A$. 
From a mesoscopic point of view, $\tilde{D}_{ii}/f_i^2$ is simply seen as a measure of fluctuations, so this may imply connections between macroscopic oscillation and mesoscopic fluctuations. 

\textit{Example of the TSL.---}
\begin{figure}
    \centering
    \includegraphics[width=\hsize]{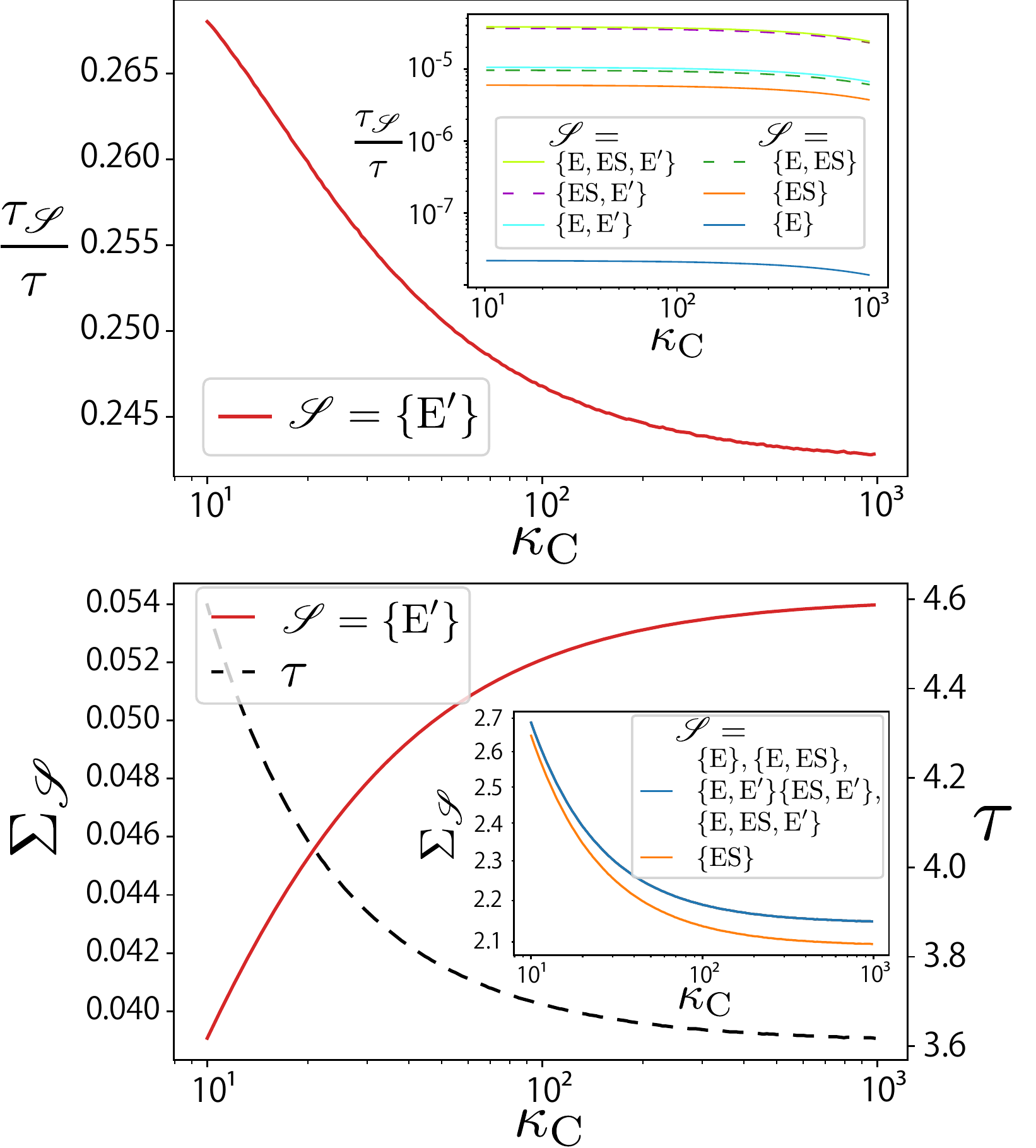}
    \caption{
    For the model of CRN~\eqref{enzyme}, we can see that the TSL holds in the shown range of parameter $\kappa_\mathrm{C}$ for all subsets (upper panel). 
    As expected from the relatively tight inequality $\tau_{\{\mathrm{E}'\}}/\tau\sim0.25<1$, there is a trade-off between speed $\tau$ and the partial dissipation $\Sigma_\mathrm{E'}$ (lower panel). 
    On the other hand, for subsets other than $\{\mathrm{E'}\}$, TSL is not a good estimation (upper inset). Therefore, the trade-off between partial dissipation and speed does not hold for them (lower inset).
    We note that we observed only a few percent of changes in the averaged diffusion coefficient and the distance between the initial and final distribution when changing the parameter $\kappa_\mr{C}$. 
    }
    \label{tsl_fig}
\end{figure}
We numerically examine the TSL and the expected trade-off relation.
To this end, we consider a model of enzymatic reaction with a coenzyme:
\begin{equation}
    \begin{split}
        \mathrm{E}+\mathrm{S}\rightleftharpoons \mathrm{ES}\rightleftharpoons \mathrm{E}+\mathrm{P},\quad
        \mathrm{E}\rightleftharpoons \mathrm{E}'+\mathrm{C}. 
    \end{split}\label{enzyme}
\end{equation}
We assume $[\mathrm{S}]$ and $[\mathrm{P}]$ are kept at constant value, and the system is first in a steady state with a certain value of $[\mathrm{C}]_0$. 
Next, the CRN comes in contact with a particle reservoir of $\mathrm{C}$, where the concentration of $\mr{C}$ is $[\mathrm{C}]^\mathrm{ext}\neq [\mathrm{C}]_0$. 
The system starts to evolve with an external flow $\mathcal{J}_\mathrm{C}^\Y=-\kappa_\mathrm{C}([\mathrm{C}]-[\mathrm{C}]^\mathrm{ext})$, where $\kappa_\mathrm{C}$ is a constant~\cite{feinberg2019foundations} (for details, see Supplemental Material~\cite{supplement}). 
We define $\tau$ as the time it takes for the system to reach another steady state.
By increasing the speed $\kappa_\mathrm{C}$ of exchanging $\mr{C}$, we can decrease $\tau$.

From the upper panel in Fig.~\ref{tsl_fig}, we can confirm the TSL. 
The TSL for $\{\mathrm{E'}\}$ gives a nice bound $\tau_\mathrm{E'}/\tau\sim 0.25$, but the other TSLs, shown in the inset, do not bound $\tau$ very well. 
As a result, while there is a clear trade-off relation between speed $\tau$ and the partial entropy production $\Sigma_\mathrm{E'}$ as we see in the lower panel in Fig.~\ref{tsl_fig}, 
the other partial entropy productions do not increase, as shown in the inset. 
Our TSL is characterized by the fact that it is possible to find a tight bound and acquire some trade-off relation by appropriately choosing a subset of chemical species.

\textit{Conclusion}.---
We have shown a TUR between the fluctuation defined by the scaled diffusion coefficient and the changing rate of concentration and dissipation, namely, the entropy production rate, in deterministic CRNs. 
We have also obtained a TSL. 
The lower bound on the time it takes when an initial concentration distribution goes to another final distribution is given by combining the entropy production and intrinsic fluctuation. 
These results are proved under quite general settings, so they reinforce the universality of TUR and TSL. 

In addition to the TSL we have derived, there exist speed limits and trade-offs in CRNs. 
One example is the information geometric speed limit~\cite{yoshimura2021information}, and its relationship with the TSL is summarized in Supplemental Material~\cite{supplement}.
Besides, many trade-off relations have been found for various biochemical processes~\cite{pfeiffer2001cooperation,lan2012energy,mehta2012energetic,govern2014energy,ito2015maxwell}. 
The relationship between these individual trade-offs and our general result of TSL is still not well understood, and future research is needed. 
We expect the general result to give a new and unified perspective to our understanding of biochemical processes. 


\begin{acknowledgments}
We thank Yoshiyuki Nakamura and Kazumasa Takeuchi for fruitful discussions. 
S.I. is supported by JSPS KAKENHI Grant No. 19H05796, 21H01560, JST Presto Grant No. JPMJPR18M2 and UTEC-UTokyo FSI Research Grant Program.
\end{acknowledgments}


\begin{thebibliography}{62}%
\makeatletter
\providecommand \@ifxundefined [1]{%
 \@ifx{#1\undefined}
}%
\providecommand \@ifnum [1]{%
 \ifnum #1\expandafter \@firstoftwo
 \else \expandafter \@secondoftwo
 \fi
}%
\providecommand \@ifx [1]{%
 \ifx #1\expandafter \@firstoftwo
 \else \expandafter \@secondoftwo
 \fi
}%
\providecommand \natexlab [1]{#1}%
\providecommand \enquote  [1]{``#1''}%
\providecommand \bibnamefont  [1]{#1}%
\providecommand \bibfnamefont [1]{#1}%
\providecommand \citenamefont [1]{#1}%
\providecommand \href@noop [0]{\@secondoftwo}%
\providecommand \href [0]{\begingroup \@sanitize@url \@href}%
\providecommand \@href[1]{\@@startlink{#1}\@@href}%
\providecommand \@@href[1]{\endgroup#1\@@endlink}%
\providecommand \@sanitize@url [0]{\catcode `\\12\catcode `\$12\catcode
  `\&12\catcode `\#12\catcode `\^12\catcode `\_12\catcode `\%12\relax}%
\providecommand \@@startlink[1]{}%
\providecommand \@@endlink[0]{}%
\providecommand \url  [0]{\begingroup\@sanitize@url \@url }%
\providecommand \@url [1]{\endgroup\@href {#1}{\urlprefix }}%
\providecommand \urlprefix  [0]{URL }%
\providecommand \Eprint [0]{\href }%
\providecommand \doibase [0]{https://doi.org/}%
\providecommand \selectlanguage [0]{\@gobble}%
\providecommand \bibinfo  [0]{\@secondoftwo}%
\providecommand \bibfield  [0]{\@secondoftwo}%
\providecommand \translation [1]{[#1]}%
\providecommand \BibitemOpen [0]{}%
\providecommand \bibitemStop [0]{}%
\providecommand \bibitemNoStop [0]{.\EOS\space}%
\providecommand \EOS [0]{\spacefactor3000\relax}%
\providecommand \BibitemShut  [1]{\csname bibitem#1\endcsname}%
\let\auto@bib@innerbib\@empty
\bibitem [{\citenamefont {Seifert}(2012)}]{seifert2012stochastic}%
  \BibitemOpen
  \bibfield  {author} {\bibinfo {author} {\bibfnamefont {U.}~\bibnamefont
  {Seifert}},\ }\href@noop {} {\bibfield  {journal} {\bibinfo  {journal} {Rep.
  Prog. Phys.}\ }\textbf {\bibinfo {volume} {75}},\ \bibinfo {pages} {126001}
  (\bibinfo {year} {2012})}\BibitemShut {NoStop}%
\bibitem [{\citenamefont {Sekimoto}(2010)}]{sekimoto2010stochastic}%
  \BibitemOpen
  \bibfield  {author} {\bibinfo {author} {\bibfnamefont {K.}~\bibnamefont
  {Sekimoto}},\ }\href@noop {} {\emph {\bibinfo {title} {Stochastic
  energetics}}}\ (\bibinfo  {publisher} {Springer},\ \bibinfo {year}
  {2010})\BibitemShut {NoStop}%
\bibitem [{\citenamefont {Barato}\ and\ \citenamefont
  {Seifert}(2015)}]{barato2015thermodynamic}%
  \BibitemOpen
  \bibfield  {author} {\bibinfo {author} {\bibfnamefont {A.~C.}\ \bibnamefont
  {Barato}}\ and\ \bibinfo {author} {\bibfnamefont {U.}~\bibnamefont
  {Seifert}},\ }\href@noop {} {\bibfield  {journal} {\bibinfo  {journal} {Phys.
  Rev. Lett.}\ }\textbf {\bibinfo {volume} {114}},\ \bibinfo {pages} {158101}
  (\bibinfo {year} {2015})}\BibitemShut {NoStop}%
\bibitem [{\citenamefont {Horowitz}\ and\ \citenamefont
  {Gingrich}(2020)}]{horowitz2020thermodynamic}%
  \BibitemOpen
  \bibfield  {author} {\bibinfo {author} {\bibfnamefont {J.~M.}\ \bibnamefont
  {Horowitz}}\ and\ \bibinfo {author} {\bibfnamefont {T.~R.}\ \bibnamefont
  {Gingrich}},\ }\href@noop {} {\bibfield  {journal} {\bibinfo  {journal} {Nat.
  Phys.}\ }\textbf {\bibinfo {volume} {16}},\ \bibinfo {pages} {15} (\bibinfo
  {year} {2020})}\BibitemShut {NoStop}%
\bibitem [{\citenamefont {Gingrich}\ \emph {et~al.}(2016)\citenamefont
  {Gingrich}, \citenamefont {Horowitz}, \citenamefont {Perunov},\ and\
  \citenamefont {England}}]{gingrich2016dissipation}%
  \BibitemOpen
  \bibfield  {author} {\bibinfo {author} {\bibfnamefont {T.~R.}\ \bibnamefont
  {Gingrich}}, \bibinfo {author} {\bibfnamefont {J.~M.}\ \bibnamefont
  {Horowitz}}, \bibinfo {author} {\bibfnamefont {N.}~\bibnamefont {Perunov}},\
  and\ \bibinfo {author} {\bibfnamefont {J.~L.}\ \bibnamefont {England}},\
  }\href@noop {} {\bibfield  {journal} {\bibinfo  {journal} {Phys. Rev. Lett.}\
  }\textbf {\bibinfo {volume} {116}},\ \bibinfo {pages} {120601} (\bibinfo
  {year} {2016})}\BibitemShut {NoStop}%
\bibitem [{\citenamefont {Pietzonka}\ \emph {et~al.}(2016)\citenamefont
  {Pietzonka}, \citenamefont {Barato},\ and\ \citenamefont
  {Seifert}}]{pietzonka2016universal}%
  \BibitemOpen
  \bibfield  {author} {\bibinfo {author} {\bibfnamefont {P.}~\bibnamefont
  {Pietzonka}}, \bibinfo {author} {\bibfnamefont {A.~C.}\ \bibnamefont
  {Barato}},\ and\ \bibinfo {author} {\bibfnamefont {U.}~\bibnamefont
  {Seifert}},\ }\href@noop {} {\bibfield  {journal} {\bibinfo  {journal} {J.
  Stat. Mech.}\ }\textbf {\bibinfo {volume} {2016}},\ \bibinfo {pages} {124004}
  (\bibinfo {year} {2016})}\BibitemShut {NoStop}%
\bibitem [{\citenamefont {Horowitz}\ and\ \citenamefont
  {Gingrich}(2017)}]{horowitz2017proof}%
  \BibitemOpen
  \bibfield  {author} {\bibinfo {author} {\bibfnamefont {J.~M.}\ \bibnamefont
  {Horowitz}}\ and\ \bibinfo {author} {\bibfnamefont {T.~R.}\ \bibnamefont
  {Gingrich}},\ }\href@noop {} {\bibfield  {journal} {\bibinfo  {journal}
  {Phys. Rev. E}\ }\textbf {\bibinfo {volume} {96}},\ \bibinfo {pages} {020103}
  (\bibinfo {year} {2017})}\BibitemShut {NoStop}%
\bibitem [{\citenamefont {Proesmans}\ and\ \citenamefont {Van~den
  Broeck}(2017)}]{proesmans2017discrete}%
  \BibitemOpen
  \bibfield  {author} {\bibinfo {author} {\bibfnamefont {K.}~\bibnamefont
  {Proesmans}}\ and\ \bibinfo {author} {\bibfnamefont {C.}~\bibnamefont
  {Van~den Broeck}},\ }\href@noop {} {\bibfield  {journal} {\bibinfo  {journal}
  {Europhys. Lett.}\ }\textbf {\bibinfo {volume} {119}},\ \bibinfo {pages}
  {20001} (\bibinfo {year} {2017})}\BibitemShut {NoStop}%
\bibitem [{\citenamefont {Pietzonka}\ \emph {et~al.}(2017)\citenamefont
  {Pietzonka}, \citenamefont {Ritort},\ and\ \citenamefont
  {Seifert}}]{pietzonka2017finite}%
  \BibitemOpen
  \bibfield  {author} {\bibinfo {author} {\bibfnamefont {P.}~\bibnamefont
  {Pietzonka}}, \bibinfo {author} {\bibfnamefont {F.}~\bibnamefont {Ritort}},\
  and\ \bibinfo {author} {\bibfnamefont {U.}~\bibnamefont {Seifert}},\
  }\href@noop {} {\bibfield  {journal} {\bibinfo  {journal} {Phys. Rev. E}\
  }\textbf {\bibinfo {volume} {96}},\ \bibinfo {pages} {012101} (\bibinfo
  {year} {2017})}\BibitemShut {NoStop}%
\bibitem [{\citenamefont {Dechant}\ and\ \citenamefont
  {Sasa}(2018)}]{dechant2018current}%
  \BibitemOpen
  \bibfield  {author} {\bibinfo {author} {\bibfnamefont {A.}~\bibnamefont
  {Dechant}}\ and\ \bibinfo {author} {\bibfnamefont {S.-i.}\ \bibnamefont
  {Sasa}},\ }\href@noop {} {\bibfield  {journal} {\bibinfo  {journal} {J. Stat.
  Mech.}\ }\textbf {\bibinfo {volume} {2018}},\ \bibinfo {pages} {063209}
  (\bibinfo {year} {2018})}\BibitemShut {NoStop}%
\bibitem [{\citenamefont {Dechant}(2018)}]{dechant2018multidimensional}%
  \BibitemOpen
  \bibfield  {author} {\bibinfo {author} {\bibfnamefont {A.}~\bibnamefont
  {Dechant}},\ }\href@noop {} {\bibfield  {journal} {\bibinfo  {journal} {J.
  Phys. A}\ }\textbf {\bibinfo {volume} {52}},\ \bibinfo {pages} {035001}
  (\bibinfo {year} {2018})}\BibitemShut {NoStop}%
\bibitem [{\citenamefont {Pietzonka}\ and\ \citenamefont
  {Seifert}(2018)}]{pietzonka2018universal}%
  \BibitemOpen
  \bibfield  {author} {\bibinfo {author} {\bibfnamefont {P.}~\bibnamefont
  {Pietzonka}}\ and\ \bibinfo {author} {\bibfnamefont {U.}~\bibnamefont
  {Seifert}},\ }\href@noop {} {\bibfield  {journal} {\bibinfo  {journal} {Phys.
  Rev. Lett.}\ }\textbf {\bibinfo {volume} {120}},\ \bibinfo {pages} {190602}
  (\bibinfo {year} {2018})}\BibitemShut {NoStop}%
\bibitem [{\citenamefont {Timpanaro}\ \emph {et~al.}(2019)\citenamefont
  {Timpanaro}, \citenamefont {Guarnieri}, \citenamefont {Goold},\ and\
  \citenamefont {Landi}}]{timpanaro2019thermodynamic}%
  \BibitemOpen
  \bibfield  {author} {\bibinfo {author} {\bibfnamefont {A.~M.}\ \bibnamefont
  {Timpanaro}}, \bibinfo {author} {\bibfnamefont {G.}~\bibnamefont
  {Guarnieri}}, \bibinfo {author} {\bibfnamefont {J.}~\bibnamefont {Goold}},\
  and\ \bibinfo {author} {\bibfnamefont {G.~T.}\ \bibnamefont {Landi}},\
  }\href@noop {} {\bibfield  {journal} {\bibinfo  {journal} {Phys. Rev. Lett.}\
  }\textbf {\bibinfo {volume} {123}},\ \bibinfo {pages} {090604} (\bibinfo
  {year} {2019})}\BibitemShut {NoStop}%
\bibitem [{\citenamefont {Hasegawa}\ and\ \citenamefont
  {Van~Vu}(2019)}]{hasegawa2019fluctuation}%
  \BibitemOpen
  \bibfield  {author} {\bibinfo {author} {\bibfnamefont {Y.}~\bibnamefont
  {Hasegawa}}\ and\ \bibinfo {author} {\bibfnamefont {T.}~\bibnamefont
  {Van~Vu}},\ }\href@noop {} {\bibfield  {journal} {\bibinfo  {journal} {Phys.
  Rev. Lett.}\ }\textbf {\bibinfo {volume} {123}},\ \bibinfo {pages} {110602}
  (\bibinfo {year} {2019})}\BibitemShut {NoStop}%
\bibitem [{\citenamefont {Falasco}\ \emph {et~al.}(2020)\citenamefont
  {Falasco}, \citenamefont {Esposito},\ and\ \citenamefont
  {Delvenne}}]{falasco2020unifying}%
  \BibitemOpen
  \bibfield  {author} {\bibinfo {author} {\bibfnamefont {G.}~\bibnamefont
  {Falasco}}, \bibinfo {author} {\bibfnamefont {M.}~\bibnamefont {Esposito}},\
  and\ \bibinfo {author} {\bibfnamefont {J.-C.}\ \bibnamefont {Delvenne}},\
  }\href@noop {} {\bibfield  {journal} {\bibinfo  {journal} {New J. Phys.}\
  }\textbf {\bibinfo {volume} {22}},\ \bibinfo {pages} {053046} (\bibinfo
  {year} {2020})}\BibitemShut {NoStop}%
\bibitem [{\citenamefont {Wolpert}(2020)}]{wolpert2020uncertainty}%
  \BibitemOpen
  \bibfield  {author} {\bibinfo {author} {\bibfnamefont {D.~H.}\ \bibnamefont
  {Wolpert}},\ }\href@noop {} {\bibfield  {journal} {\bibinfo  {journal} {Phys.
  Rev. Lett.}\ }\textbf {\bibinfo {volume} {125}},\ \bibinfo {pages} {200602}
  (\bibinfo {year} {2020})}\BibitemShut {NoStop}%
\bibitem [{\citenamefont {Otsubo}\ \emph {et~al.}(2020)\citenamefont {Otsubo},
  \citenamefont {Ito}, \citenamefont {Dechant},\ and\ \citenamefont
  {Sagawa}}]{otsubo2020estimating}%
  \BibitemOpen
  \bibfield  {author} {\bibinfo {author} {\bibfnamefont {S.}~\bibnamefont
  {Otsubo}}, \bibinfo {author} {\bibfnamefont {S.}~\bibnamefont {Ito}},
  \bibinfo {author} {\bibfnamefont {A.}~\bibnamefont {Dechant}},\ and\ \bibinfo
  {author} {\bibfnamefont {T.}~\bibnamefont {Sagawa}},\ }\href@noop {}
  {\bibfield  {journal} {\bibinfo  {journal} {Phys. Rev. E}\ }\textbf {\bibinfo
  {volume} {101}},\ \bibinfo {pages} {062106} (\bibinfo {year}
  {2020})}\BibitemShut {NoStop}%
\bibitem [{\citenamefont {Manikandan}\ \emph {et~al.}(2020)\citenamefont
  {Manikandan}, \citenamefont {Gupta},\ and\ \citenamefont
  {Krishnamurthy}}]{manikandan2020inferring}%
  \BibitemOpen
  \bibfield  {author} {\bibinfo {author} {\bibfnamefont {S.~K.}\ \bibnamefont
  {Manikandan}}, \bibinfo {author} {\bibfnamefont {D.}~\bibnamefont {Gupta}},\
  and\ \bibinfo {author} {\bibfnamefont {S.}~\bibnamefont {Krishnamurthy}},\
  }\href@noop {} {\bibfield  {journal} {\bibinfo  {journal} {Phys. Rev. Lett.}\
  }\textbf {\bibinfo {volume} {124}},\ \bibinfo {pages} {120603} (\bibinfo
  {year} {2020})}\BibitemShut {NoStop}%
\bibitem [{\citenamefont {Liu}\ \emph {et~al.}(2020)\citenamefont {Liu},
  \citenamefont {Gong},\ and\ \citenamefont {Ueda}}]{liu2020thermodynamic}%
  \BibitemOpen
  \bibfield  {author} {\bibinfo {author} {\bibfnamefont {K.}~\bibnamefont
  {Liu}}, \bibinfo {author} {\bibfnamefont {Z.}~\bibnamefont {Gong}},\ and\
  \bibinfo {author} {\bibfnamefont {M.}~\bibnamefont {Ueda}},\ }\href@noop {}
  {\bibfield  {journal} {\bibinfo  {journal} {Phys. Rev. Lett.}\ }\textbf
  {\bibinfo {volume} {125}},\ \bibinfo {pages} {140602} (\bibinfo {year}
  {2020})}\BibitemShut {NoStop}%
\bibitem [{\citenamefont {Hasegawa}(2021)}]{hasegawa2021thermodynamic}%
  \BibitemOpen
  \bibfield  {author} {\bibinfo {author} {\bibfnamefont {Y.}~\bibnamefont
  {Hasegawa}},\ }\href@noop {} {\bibfield  {journal} {\bibinfo  {journal}
  {Phys. Rev. Lett.}\ }\textbf {\bibinfo {volume} {126}},\ \bibinfo {pages}
  {010602} (\bibinfo {year} {2021})}\BibitemShut {NoStop}%
\bibitem [{\citenamefont {Shiraishi}\ \emph {et~al.}(2018)\citenamefont
  {Shiraishi}, \citenamefont {Funo},\ and\ \citenamefont
  {Saito}}]{shiraishi2018speed}%
  \BibitemOpen
  \bibfield  {author} {\bibinfo {author} {\bibfnamefont {N.}~\bibnamefont
  {Shiraishi}}, \bibinfo {author} {\bibfnamefont {K.}~\bibnamefont {Funo}},\
  and\ \bibinfo {author} {\bibfnamefont {K.}~\bibnamefont {Saito}},\
  }\href@noop {} {\bibfield  {journal} {\bibinfo  {journal} {Phys. Rev. Lett.}\
  }\textbf {\bibinfo {volume} {121}},\ \bibinfo {pages} {070601} (\bibinfo
  {year} {2018})}\BibitemShut {NoStop}%
\bibitem [{\citenamefont {Funo}\ \emph {et~al.}(2019)\citenamefont {Funo},
  \citenamefont {Shiraishi},\ and\ \citenamefont {Saito}}]{funo2019speed}%
  \BibitemOpen
  \bibfield  {author} {\bibinfo {author} {\bibfnamefont {K.}~\bibnamefont
  {Funo}}, \bibinfo {author} {\bibfnamefont {N.}~\bibnamefont {Shiraishi}},\
  and\ \bibinfo {author} {\bibfnamefont {K.}~\bibnamefont {Saito}},\
  }\href@noop {} {\bibfield  {journal} {\bibinfo  {journal} {New J. Phys.}\
  }\textbf {\bibinfo {volume} {21}},\ \bibinfo {pages} {013006} (\bibinfo
  {year} {2019})}\BibitemShut {NoStop}%
\bibitem [{\citenamefont {Van~Vu}\ \emph {et~al.}(2020)\citenamefont {Van~Vu},
  \citenamefont {Hasegawa} \emph {et~al.}}]{van2020unified}%
  \BibitemOpen
  \bibfield  {author} {\bibinfo {author} {\bibfnamefont {T.}~\bibnamefont
  {Van~Vu}}, \bibinfo {author} {\bibfnamefont {Y.}~\bibnamefont {Hasegawa}},
  \emph {et~al.},\ }\href@noop {} {\bibfield  {journal} {\bibinfo  {journal}
  {Phys. Rev. E}\ }\textbf {\bibinfo {volume} {102}},\ \bibinfo {pages}
  {062132} (\bibinfo {year} {2020})}\BibitemShut {NoStop}%
\bibitem [{\citenamefont {Falasco}\ and\ \citenamefont
  {Esposito}(2020)}]{falasco2020dissipation}%
  \BibitemOpen
  \bibfield  {author} {\bibinfo {author} {\bibfnamefont {G.}~\bibnamefont
  {Falasco}}\ and\ \bibinfo {author} {\bibfnamefont {M.}~\bibnamefont
  {Esposito}},\ }\href@noop {} {\bibfield  {journal} {\bibinfo  {journal}
  {Phys. Rev. Lett.}\ }\textbf {\bibinfo {volume} {125}},\ \bibinfo {pages}
  {120604} (\bibinfo {year} {2020})}\BibitemShut {NoStop}%
\bibitem [{\citenamefont {Mandelstam}\ and\ \citenamefont
  {Tamm}(1945)}]{mandelstam1945uncertainty}%
  \BibitemOpen
  \bibfield  {author} {\bibinfo {author} {\bibfnamefont {L.}~\bibnamefont
  {Mandelstam}}\ and\ \bibinfo {author} {\bibfnamefont {I.}~\bibnamefont
  {Tamm}},\ }\href@noop {} {\bibfield  {journal} {\bibinfo  {journal} {J. Phys.
  USSR}\ }\textbf {\bibinfo {volume} {9}},\ \bibinfo {pages} {249} (\bibinfo
  {year} {1945})}\BibitemShut {NoStop}%
\bibitem [{\citenamefont {Margolus}\ and\ \citenamefont
  {Levitin}(1998)}]{margolus1998maximum}%
  \BibitemOpen
  \bibfield  {author} {\bibinfo {author} {\bibfnamefont {N.}~\bibnamefont
  {Margolus}}\ and\ \bibinfo {author} {\bibfnamefont {L.~B.}\ \bibnamefont
  {Levitin}},\ }\href@noop {} {\bibfield  {journal} {\bibinfo  {journal}
  {Physica D: Nonlinear Phenomena}\ }\textbf {\bibinfo {volume} {120}},\
  \bibinfo {pages} {188} (\bibinfo {year} {1998})}\BibitemShut {NoStop}%
\bibitem [{\citenamefont {Aurell}\ \emph {et~al.}(2012)\citenamefont {Aurell},
  \citenamefont {Gaw\c{e}dzki}, \citenamefont {Mej\'{\i}a-Monasterio},
  \citenamefont {Mohayaee},\ and\ \citenamefont
  {Muratore-Ginanneschi}}]{aurell2012refined}%
  \BibitemOpen
  \bibfield  {author} {\bibinfo {author} {\bibfnamefont {E.}~\bibnamefont
  {Aurell}}, \bibinfo {author} {\bibfnamefont {K.}~\bibnamefont
  {Gaw\c{e}dzki}}, \bibinfo {author} {\bibfnamefont {C.}~\bibnamefont
  {Mej\'{\i}a-Monasterio}}, \bibinfo {author} {\bibfnamefont {R.}~\bibnamefont
  {Mohayaee}},\ and\ \bibinfo {author} {\bibfnamefont {P.}~\bibnamefont
  {Muratore-Ginanneschi}},\ }\href@noop {} {\bibfield  {journal} {\bibinfo
  {journal} {J. Stat. Phys.}\ }\textbf {\bibinfo {volume} {147}} (\bibinfo
  {year} {2012})}\BibitemShut {NoStop}%
\bibitem [{\citenamefont {Pires}\ \emph {et~al.}(2016)\citenamefont {Pires},
  \citenamefont {Cianciaruso}, \citenamefont {C{\'e}leri}, \citenamefont
  {Adesso},\ and\ \citenamefont {Soares-Pinto}}]{pires2016generalized}%
  \BibitemOpen
  \bibfield  {author} {\bibinfo {author} {\bibfnamefont {D.~P.}\ \bibnamefont
  {Pires}}, \bibinfo {author} {\bibfnamefont {M.}~\bibnamefont {Cianciaruso}},
  \bibinfo {author} {\bibfnamefont {L.~C.}\ \bibnamefont {C{\'e}leri}},
  \bibinfo {author} {\bibfnamefont {G.}~\bibnamefont {Adesso}},\ and\ \bibinfo
  {author} {\bibfnamefont {D.~O.}\ \bibnamefont {Soares-Pinto}},\ }\href@noop
  {} {\bibfield  {journal} {\bibinfo  {journal} {Phys. Rev. X}\ }\textbf
  {\bibinfo {volume} {6}},\ \bibinfo {pages} {021031} (\bibinfo {year}
  {2016})}\BibitemShut {NoStop}%
\bibitem [{\citenamefont {Okuyama}\ and\ \citenamefont
  {Ohzeki}(2018)}]{okuyama2018quantum}%
  \BibitemOpen
  \bibfield  {author} {\bibinfo {author} {\bibfnamefont {M.}~\bibnamefont
  {Okuyama}}\ and\ \bibinfo {author} {\bibfnamefont {M.}~\bibnamefont
  {Ohzeki}},\ }\href@noop {} {\bibfield  {journal} {\bibinfo  {journal} {Phys.
  Rev. Lett.}\ }\textbf {\bibinfo {volume} {120}},\ \bibinfo {pages} {070402}
  (\bibinfo {year} {2018})}\BibitemShut {NoStop}%
\bibitem [{\citenamefont {Ito}(2018)}]{ito2018stochastic}%
  \BibitemOpen
  \bibfield  {author} {\bibinfo {author} {\bibfnamefont {S.}~\bibnamefont
  {Ito}},\ }\href@noop {} {\bibfield  {journal} {\bibinfo  {journal} {Phys.
  Rev. Lett.}\ }\textbf {\bibinfo {volume} {121}},\ \bibinfo {pages} {030605}
  (\bibinfo {year} {2018})}\BibitemShut {NoStop}%
\bibitem [{\citenamefont {Dechant}\ and\ \citenamefont
  {Sakurai}(2019)}]{dechant2019thermodynamic}%
  \BibitemOpen
  \bibfield  {author} {\bibinfo {author} {\bibfnamefont {A.}~\bibnamefont
  {Dechant}}\ and\ \bibinfo {author} {\bibfnamefont {Y.}~\bibnamefont
  {Sakurai}},\ }\Eprint {https://arxiv.org/abs/1912.08405} {arXiv:1912.08405}
  (\bibinfo {year} {2019})\BibitemShut {NoStop}%
\bibitem [{\citenamefont {Ito}\ and\ \citenamefont
  {Dechant}(2020)}]{ito2020stochastic}%
  \BibitemOpen
  \bibfield  {author} {\bibinfo {author} {\bibfnamefont {S.}~\bibnamefont
  {Ito}}\ and\ \bibinfo {author} {\bibfnamefont {A.}~\bibnamefont {Dechant}},\
  }\href@noop {} {\bibfield  {journal} {\bibinfo  {journal} {Phys. Rev. X}\
  }\textbf {\bibinfo {volume} {10}},\ \bibinfo {pages} {021056} (\bibinfo
  {year} {2020})}\BibitemShut {NoStop}%
\bibitem [{\citenamefont {Gupta}\ and\ \citenamefont
  {Busiello}(2020)}]{gupta2020tighter}%
  \BibitemOpen
  \bibfield  {author} {\bibinfo {author} {\bibfnamefont {D.}~\bibnamefont
  {Gupta}}\ and\ \bibinfo {author} {\bibfnamefont {D.~M.}\ \bibnamefont
  {Busiello}},\ }\href@noop {} {\bibfield  {journal} {\bibinfo  {journal}
  {Phys. Rev. E}\ }\textbf {\bibinfo {volume} {102}},\ \bibinfo {pages}
  {062121} (\bibinfo {year} {2020})}\BibitemShut {NoStop}%
\bibitem [{\citenamefont {Nicholson}\ \emph {et~al.}(2020)\citenamefont
  {Nicholson}, \citenamefont {Garcia-Pintos}, \citenamefont {del Campo},\ and\
  \citenamefont {Green}}]{nicholson2020time}%
  \BibitemOpen
  \bibfield  {author} {\bibinfo {author} {\bibfnamefont {S.~B.}\ \bibnamefont
  {Nicholson}}, \bibinfo {author} {\bibfnamefont {L.~P.}\ \bibnamefont
  {Garcia-Pintos}}, \bibinfo {author} {\bibfnamefont {A.}~\bibnamefont {del
  Campo}},\ and\ \bibinfo {author} {\bibfnamefont {J.~R.}\ \bibnamefont
  {Green}},\ }\href@noop {} {\bibfield  {journal} {\bibinfo  {journal} {Nat.
  Phys.}\ }\textbf {\bibinfo {volume} {16}},\ \bibinfo {pages} {1211} (\bibinfo
  {year} {2020})}\BibitemShut {NoStop}%
\bibitem [{\citenamefont {Yoshimura}\ and\ \citenamefont
  {Ito}(2021)}]{yoshimura2021information}%
  \BibitemOpen
  \bibfield  {author} {\bibinfo {author} {\bibfnamefont {K.}~\bibnamefont
  {Yoshimura}}\ and\ \bibinfo {author} {\bibfnamefont {S.}~\bibnamefont
  {Ito}},\ }\href@noop {} {\bibfield  {journal} {\bibinfo  {journal} {Phys.
  Rev. Research}\ }\textbf {\bibinfo {volume} {3}},\ \bibinfo {pages} {013175}
  (\bibinfo {year} {2021})}\BibitemShut {NoStop}%
\bibitem [{\citenamefont {Van~Vu}\ and\ \citenamefont
  {Hasegawa}(2021)}]{van2021geometrical}%
  \BibitemOpen
  \bibfield  {author} {\bibinfo {author} {\bibfnamefont {T.}~\bibnamefont
  {Van~Vu}}\ and\ \bibinfo {author} {\bibfnamefont {Y.}~\bibnamefont
  {Hasegawa}},\ }\href@noop {} {\bibfield  {journal} {\bibinfo  {journal}
  {Phys. Rev. Lett.}\ }\textbf {\bibinfo {volume} {126}},\ \bibinfo {pages}
  {010601} (\bibinfo {year} {2021})}\BibitemShut {NoStop}%
\bibitem [{\citenamefont {Nakazato}\ and\ \citenamefont
  {Ito}(2021)}]{nakazato2021geometrical}%
  \BibitemOpen
  \bibfield  {author} {\bibinfo {author} {\bibfnamefont {M.}~\bibnamefont
  {Nakazato}}\ and\ \bibinfo {author} {\bibfnamefont {S.}~\bibnamefont {Ito}},\
  }\Eprint {https://arxiv.org/abs/2103.00503} {arXiv:2103.00503}  (\bibinfo
  {year} {2021})\BibitemShut {NoStop}%
\bibitem [{\citenamefont {Gibbs}(1878)}]{gibbs1878equilibrium}%
  \BibitemOpen
  \bibfield  {author} {\bibinfo {author} {\bibfnamefont {J.~W.}\ \bibnamefont
  {Gibbs}},\ }\href@noop {} {\bibfield  {journal} {\bibinfo  {journal} {Am. J.
  Sci.}\ }\textbf {\bibinfo {volume} {s3-16}},\ \bibinfo {pages} {441}
  (\bibinfo {year} {1878})}\BibitemShut {NoStop}%
\bibitem [{\citenamefont {De~Donder}\ and\ \citenamefont
  {Van~Rysselberghe}(1936)}]{de1936thermodynamic}%
  \BibitemOpen
  \bibfield  {author} {\bibinfo {author} {\bibfnamefont {T.}~\bibnamefont
  {De~Donder}}\ and\ \bibinfo {author} {\bibfnamefont {P.}~\bibnamefont
  {Van~Rysselberghe}},\ }\href@noop {} {\emph {\bibinfo {title} {Thermodynamic
  theory of affinity: A book of principles}}},\ Vol.~\bibinfo {volume} {1}\
  (\bibinfo  {publisher} {Stanford university press},\ \bibinfo {year}
  {1936})\BibitemShut {NoStop}%
\bibitem [{\citenamefont {Kondepudi}\ and\ \citenamefont
  {Prigogine}(2014)}]{kondepudi2014modern}%
  \BibitemOpen
  \bibfield  {author} {\bibinfo {author} {\bibfnamefont {D.}~\bibnamefont
  {Kondepudi}}\ and\ \bibinfo {author} {\bibfnamefont {I.}~\bibnamefont
  {Prigogine}},\ }\href@noop {} {\emph {\bibinfo {title} {Modern
  thermodynamics: from heat engines to dissipative structures}}}\ (\bibinfo
  {publisher} {John Wiley \& Sons},\ \bibinfo {year} {2014})\BibitemShut
  {NoStop}%
\bibitem [{\citenamefont {Ge}\ and\ \citenamefont
  {Qian}(2016{\natexlab{a}})}]{ge2016mesoscopic}%
  \BibitemOpen
  \bibfield  {author} {\bibinfo {author} {\bibfnamefont {H.}~\bibnamefont
  {Ge}}\ and\ \bibinfo {author} {\bibfnamefont {H.}~\bibnamefont {Qian}},\
  }\href@noop {} {\bibfield  {journal} {\bibinfo  {journal} {Phys. Rev. E}\
  }\textbf {\bibinfo {volume} {94}},\ \bibinfo {pages} {052150} (\bibinfo
  {year} {2016}{\natexlab{a}})}\BibitemShut {NoStop}%
\bibitem [{\citenamefont {Ge}\ and\ \citenamefont
  {Qian}(2017)}]{ge2017mathematical}%
  \BibitemOpen
  \bibfield  {author} {\bibinfo {author} {\bibfnamefont {H.}~\bibnamefont
  {Ge}}\ and\ \bibinfo {author} {\bibfnamefont {H.}~\bibnamefont {Qian}},\
  }\href@noop {} {\bibfield  {journal} {\bibinfo  {journal} {J. Stat. Phys.}\
  }\textbf {\bibinfo {volume} {166}},\ \bibinfo {pages} {190} (\bibinfo {year}
  {2017})}\BibitemShut {NoStop}%
\bibitem [{\citenamefont {Beard}\ and\ \citenamefont
  {Qian}(2008)}]{beard2008chemical}%
  \BibitemOpen
  \bibfield  {author} {\bibinfo {author} {\bibfnamefont {D.~A.}\ \bibnamefont
  {Beard}}\ and\ \bibinfo {author} {\bibfnamefont {H.}~\bibnamefont {Qian}},\
  }\href@noop {} {\emph {\bibinfo {title} {Chemical biophysics: quantitative
  analysis of cellular systems}}},\ Vol.\ \bibinfo {volume} {126}\ (\bibinfo
  {publisher} {Cambridge University Press Cambridge},\ \bibinfo {year}
  {2008})\BibitemShut {NoStop}%
\bibitem [{\citenamefont {Kurtz}(1972)}]{kurtz1972relationship}%
  \BibitemOpen
  \bibfield  {author} {\bibinfo {author} {\bibfnamefont {T.~G.}\ \bibnamefont
  {Kurtz}},\ }\href@noop {} {\bibfield  {journal} {\bibinfo  {journal} {J.
  Chem. Phys.}\ }\textbf {\bibinfo {volume} {57}},\ \bibinfo {pages} {2976}
  (\bibinfo {year} {1972})}\BibitemShut {NoStop}%
\bibitem [{\citenamefont {Kurtz}(1978)}]{kurtz1978strong}%
  \BibitemOpen
  \bibfield  {author} {\bibinfo {author} {\bibfnamefont {T.~G.}\ \bibnamefont
  {Kurtz}},\ }\href@noop {} {\bibfield  {journal} {\bibinfo  {journal} {Stoch.
  Proc. Appl.}\ }\textbf {\bibinfo {volume} {6}},\ \bibinfo {pages} {223}
  (\bibinfo {year} {1978})}\BibitemShut {NoStop}%
\bibitem [{\citenamefont {Ge}\ and\ \citenamefont
  {Qian}(2016{\natexlab{b}})}]{ge2016nonequilibrium}%
  \BibitemOpen
  \bibfield  {author} {\bibinfo {author} {\bibfnamefont {H.}~\bibnamefont
  {Ge}}\ and\ \bibinfo {author} {\bibfnamefont {H.}~\bibnamefont {Qian}},\
  }\href@noop {} {\bibfield  {journal} {\bibinfo  {journal} {Chem. Phys.}\
  }\textbf {\bibinfo {volume} {472}},\ \bibinfo {pages} {241} (\bibinfo {year}
  {2016}{\natexlab{b}})}\BibitemShut {NoStop}%
\bibitem [{\citenamefont {Rao}\ and\ \citenamefont
  {Esposito}(2016)}]{rao2016nonequilibrium}%
  \BibitemOpen
  \bibfield  {author} {\bibinfo {author} {\bibfnamefont {R.}~\bibnamefont
  {Rao}}\ and\ \bibinfo {author} {\bibfnamefont {M.}~\bibnamefont {Esposito}},\
  }\href@noop {} {\bibfield  {journal} {\bibinfo  {journal} {Phys. Rev. X}\
  }\textbf {\bibinfo {volume} {6}},\ \bibinfo {pages} {041064} (\bibinfo {year}
  {2016})}\BibitemShut {NoStop}%
\bibitem [{\citenamefont {Falasco}\ \emph {et~al.}(2018)\citenamefont
  {Falasco}, \citenamefont {Rao},\ and\ \citenamefont
  {Esposito}}]{falasco2018information}%
  \BibitemOpen
  \bibfield  {author} {\bibinfo {author} {\bibfnamefont {G.}~\bibnamefont
  {Falasco}}, \bibinfo {author} {\bibfnamefont {R.}~\bibnamefont {Rao}},\ and\
  \bibinfo {author} {\bibfnamefont {M.}~\bibnamefont {Esposito}},\ }\href@noop
  {} {\bibfield  {journal} {\bibinfo  {journal} {Phys. Rev. Lett.}\ }\textbf
  {\bibinfo {volume} {121}},\ \bibinfo {pages} {108301} (\bibinfo {year}
  {2018})}\BibitemShut {NoStop}%
\bibitem [{\citenamefont {Wachtel}\ \emph {et~al.}(2018)\citenamefont
  {Wachtel}, \citenamefont {Rao},\ and\ \citenamefont
  {Esposito}}]{wachtel2018thermodynamically}%
  \BibitemOpen
  \bibfield  {author} {\bibinfo {author} {\bibfnamefont {A.}~\bibnamefont
  {Wachtel}}, \bibinfo {author} {\bibfnamefont {R.}~\bibnamefont {Rao}},\ and\
  \bibinfo {author} {\bibfnamefont {M.}~\bibnamefont {Esposito}},\ }\href@noop
  {} {\bibfield  {journal} {\bibinfo  {journal} {New J. Phys.}\ }\textbf
  {\bibinfo {volume} {20}},\ \bibinfo {pages} {042002} (\bibinfo {year}
  {2018})}\BibitemShut {NoStop}%
\bibitem [{\citenamefont {Avanzini}\ \emph {et~al.}(2020)\citenamefont
  {Avanzini}, \citenamefont {Falasco},\ and\ \citenamefont
  {Esposito}}]{avanzini2020thermodynamics}%
  \BibitemOpen
  \bibfield  {author} {\bibinfo {author} {\bibfnamefont {F.}~\bibnamefont
  {Avanzini}}, \bibinfo {author} {\bibfnamefont {G.}~\bibnamefont {Falasco}},\
  and\ \bibinfo {author} {\bibfnamefont {M.}~\bibnamefont {Esposito}},\
  }\href@noop {} {\bibfield  {journal} {\bibinfo  {journal} {New J. Phys.}\
  }\textbf {\bibinfo {volume} {22}},\ \bibinfo {pages} {093040} (\bibinfo
  {year} {2020})}\BibitemShut {NoStop}%
\bibitem [{\citenamefont {Avanzini}\ \emph {et~al.}(2021)\citenamefont
  {Avanzini}, \citenamefont {Penocchio}, \citenamefont {Falasco},\ and\
  \citenamefont {Esposito}}]{avanzini2021nonequilibrium}%
  \BibitemOpen
  \bibfield  {author} {\bibinfo {author} {\bibfnamefont {F.}~\bibnamefont
  {Avanzini}}, \bibinfo {author} {\bibfnamefont {E.}~\bibnamefont {Penocchio}},
  \bibinfo {author} {\bibfnamefont {G.}~\bibnamefont {Falasco}},\ and\ \bibinfo
  {author} {\bibfnamefont {M.}~\bibnamefont {Esposito}},\ }\href@noop {}
  {\bibfield  {journal} {\bibinfo  {journal} {J. Chem. Phys.}\ }\textbf
  {\bibinfo {volume} {154}},\ \bibinfo {pages} {094114} (\bibinfo {year}
  {2021})}\BibitemShut {NoStop}%
\bibitem [{\citenamefont {Schnakenberg}(1976)}]{schnakenberg1976network}%
  \BibitemOpen
  \bibfield  {author} {\bibinfo {author} {\bibfnamefont {J.}~\bibnamefont
  {Schnakenberg}},\ }\href@noop {} {\bibfield  {journal} {\bibinfo  {journal}
  {Rev. Mod. Phys.}\ }\textbf {\bibinfo {volume} {48}},\ \bibinfo {pages} {571}
  (\bibinfo {year} {1976})}\BibitemShut {NoStop}%
\bibitem [{\citenamefont {Gillespie}(1992)}]{gillespie1992rigorous}%
  \BibitemOpen
  \bibfield  {author} {\bibinfo {author} {\bibfnamefont {D.~T.}\ \bibnamefont
  {Gillespie}},\ }\href@noop {} {\bibfield  {journal} {\bibinfo  {journal}
  {Physica A}\ }\textbf {\bibinfo {volume} {188}},\ \bibinfo {pages} {404}
  (\bibinfo {year} {1992})}\BibitemShut {NoStop}%
\bibitem [{\citenamefont {Gillespie}(2000)}]{gillespie2000chemical}%
  \BibitemOpen
  \bibfield  {author} {\bibinfo {author} {\bibfnamefont {D.~T.}\ \bibnamefont
  {Gillespie}},\ }\href@noop {} {\bibfield  {journal} {\bibinfo  {journal} {J.
  Chem. Phys.}\ }\textbf {\bibinfo {volume} {113}},\ \bibinfo {pages} {297}
  (\bibinfo {year} {2000})}\BibitemShut {NoStop}%
\bibitem{supplement} Supplemental Material [url], which includes Ref.~\cite{gardiner2009stochastic}.
\bibitem [{\citenamefont {Strogatz}(2018)}]{strogatz2018nonlinear}%
  \BibitemOpen
  \bibfield  {author} {\bibinfo {author} {\bibfnamefont {S.~H.}\ \bibnamefont
  {Strogatz}},\ }\href@noop {} {\emph {\bibinfo {title} {Nonlinear dynamics and
  chaos with student solutions manual: With applications to physics, biology,
  chemistry, and engineering}}}\ (\bibinfo  {publisher} {CRC press},\ \bibinfo
  {year} {2018})\BibitemShut {NoStop}%
\bibitem [{\citenamefont {Feinberg}(2019)}]{feinberg2019foundations}%
  \BibitemOpen
  \bibfield  {author} {\bibinfo {author} {\bibfnamefont {M.}~\bibnamefont
  {Feinberg}},\ }\href@noop {} {\emph {\bibinfo {title} {Foundations of
  chemical reaction network theory}}}\ (\bibinfo  {publisher} {Springer},\
  \bibinfo {year} {2019})\BibitemShut {NoStop}%
\bibitem [{\citenamefont {Pfeiffer}\ \emph {et~al.}(2001)\citenamefont
  {Pfeiffer}, \citenamefont {Schuster},\ and\ \citenamefont
  {Bonhoeffer}}]{pfeiffer2001cooperation}%
  \BibitemOpen
  \bibfield  {author} {\bibinfo {author} {\bibfnamefont {T.}~\bibnamefont
  {Pfeiffer}}, \bibinfo {author} {\bibfnamefont {S.}~\bibnamefont {Schuster}},\
  and\ \bibinfo {author} {\bibfnamefont {S.}~\bibnamefont {Bonhoeffer}},\
  }\href@noop {} {\bibfield  {journal} {\bibinfo  {journal} {Science}\ }\textbf
  {\bibinfo {volume} {292}},\ \bibinfo {pages} {504} (\bibinfo {year}
  {2001})}\BibitemShut {NoStop}%
\bibitem [{\citenamefont {Lan}\ \emph {et~al.}(2012)\citenamefont {Lan},
  \citenamefont {Sartori}, \citenamefont {Neumann}, \citenamefont {Sourjik},\
  and\ \citenamefont {Tu}}]{lan2012energy}%
  \BibitemOpen
  \bibfield  {author} {\bibinfo {author} {\bibfnamefont {G.}~\bibnamefont
  {Lan}}, \bibinfo {author} {\bibfnamefont {P.}~\bibnamefont {Sartori}},
  \bibinfo {author} {\bibfnamefont {S.}~\bibnamefont {Neumann}}, \bibinfo
  {author} {\bibfnamefont {V.}~\bibnamefont {Sourjik}},\ and\ \bibinfo {author}
  {\bibfnamefont {Y.}~\bibnamefont {Tu}},\ }\href@noop {} {\bibfield  {journal}
  {\bibinfo  {journal} {Nature physics}\ }\textbf {\bibinfo {volume} {8}},\
  \bibinfo {pages} {422} (\bibinfo {year} {2012})}\BibitemShut {NoStop}%
\bibitem [{\citenamefont {Mehta}\ and\ \citenamefont
  {Schwab}(2012)}]{mehta2012energetic}%
  \BibitemOpen
  \bibfield  {author} {\bibinfo {author} {\bibfnamefont {P.}~\bibnamefont
  {Mehta}}\ and\ \bibinfo {author} {\bibfnamefont {D.~J.}\ \bibnamefont
  {Schwab}},\ }\href@noop {} {\bibfield  {journal} {\bibinfo  {journal}
  {Proceedings of the National Academy of Sciences}\ }\textbf {\bibinfo
  {volume} {109}},\ \bibinfo {pages} {17978} (\bibinfo {year}
  {2012})}\BibitemShut {NoStop}%
\bibitem [{\citenamefont {Govern}\ and\ \citenamefont {ten
  Wolde}(2014)}]{govern2014energy}%
  \BibitemOpen
  \bibfield  {author} {\bibinfo {author} {\bibfnamefont {C.~C.}\ \bibnamefont
  {Govern}}\ and\ \bibinfo {author} {\bibfnamefont {P.~R.}\ \bibnamefont {ten
  Wolde}},\ }\href@noop {} {\bibfield  {journal} {\bibinfo  {journal} {Physical
  review letters}\ }\textbf {\bibinfo {volume} {113}},\ \bibinfo {pages}
  {258102} (\bibinfo {year} {2014})}\BibitemShut {NoStop}%
\bibitem [{\citenamefont {Ito}\ and\ \citenamefont
  {Sagawa}(2015)}]{ito2015maxwell}%
  \BibitemOpen
  \bibfield  {author} {\bibinfo {author} {\bibfnamefont {S.}~\bibnamefont
  {Ito}}\ and\ \bibinfo {author} {\bibfnamefont {T.}~\bibnamefont {Sagawa}},\
  }\href@noop {} {\bibfield  {journal} {\bibinfo  {journal} {Nature
  communications}\ }\textbf {\bibinfo {volume} {6}},\ \bibinfo {pages} {1}
  (\bibinfo {year} {2015})}\BibitemShut {NoStop}%
\bibitem [{\citenamefont {Gardiner}(2009)}]{gardiner2009stochastic}%
  \BibitemOpen
  \bibfield  {author} {\bibinfo {author} {\bibfnamefont {C.}~\bibnamefont
  {Gardiner}},\ }\href@noop {} {\emph {\bibinfo {title} {Stochastic
  methods}}},\ Vol.~\bibinfo {volume} {4}\ (\bibinfo  {publisher} {Springer
  Berlin},\ \bibinfo {year} {2009})\BibitemShut {NoStop}%
\end{thebibliography}
%

\newpage
\onecolumngrid
\appendix

\begin{center}
    \bigskip
    {\large\textbf{Supplemental Material}}
\end{center}

\section{Derivation of the Fokker--Planck equation~\eqref{fpeq} from the master equation}
Here, we derive the Fokker--Planck equation~\eqref{fpeq} from the master equation of chemical reactions. 
We introduce a notation that is useful in stochastic description. 
We label the forward and backward reactions separately unlike the main text. We assign each $\rho\in\mathscr{R}$ to the forward reaction of the so-called $\rho$th reaction in the main text and let $\overline{\rho}=\rho+M$ designate the pair backward reaction. We define $\bar{\mathscr{R}}:=\{M+1,\dots,2M\}$. Let $\mathsf{S}_\rho:=(\mathsf{S}_{1\rho},\dots,\mathsf{S}_{N\rho})^\mathrm{T}$ and $\st_{\bar{\rho}}:=-\st_{\rho}$ for each $\rho\in\mathscr{R}$. 

We consider a stochastic process where the reactions randomly occur and the number of molecules $\bm{n}=(n_1,\dots,n_N)^\mathrm{T}$ is described by the master equation: 
\begin{align}
    {\pdv{p}{t}}(t,\bm{n})
    =\sum_{\rho\in\mathscr{R}\cup\bar{\mathscr{R}}}\bqty{
    w_\rho(\bm{n}-\st_{\rho})p(t,\bm{n}-\st_{\rho})
    -w_\rho(\bm{n})p(t,\bm{n})
    }, \label{mastereq}
\end{align}
where $p(t,\bm{n})$ is the probability that the number of molecules is $\bm{n}$ at time $t$ and $w_\rho(\bm{n})$ is the occurrence rate of the $\rho$th reaction. 

Then, we consider the expansion of the right hand side of Eq.~\eqref{mastereq} by the volume $V$. Let $\bm{x}=\bm{n}/V$, $V^{-N}\tilde{p}(t,\bm{x})=p(t,V\bm{x})$, and $\tilde{w}_\rho(\bm{x})=w_\rho(V\bm{x})$. 
The function $\tilde{p}$ becomes the parbability density function when we take the limit of $V\to\infty$.
By doing the Taylor expansion of the right hand side of Eq.~\eqref{mastereq} with respect to $\st_{\rho}/V$, we have 
\begin{align}
    &\tilde{w}_\rho(\bm{x}-\st_{\rho}/V)\tilde{p}(t,\bm{x}-\st_{\rho}/V)
    -\tilde{w}_\rho(\bm{x})\tilde{p}(t,\bm{x})\\
    =&
    \sum_{k=1}^\infty
    \underset{m_1+\cdots+m_N=k}{\sum_{m_1,\dots,m_N\geq 0}}
    \bqty{\prod_{i\in\mathscr{S}_\X}\frac{1}{m_i!}
    \qty(-\frac{\st_{i\rho}}{V})^{m_i}\frac{\partial^{m_i}}{\partial x_i^{m_i}}}
    [\tilde{w}_\rho(\bm{x})\tilde{p}(t,\bm{x})].
    \label{sse}
\end{align}
Because $\tilde{w}_\rho(\bm{x})=w_\rho(V\bm{x})=\mathcal{O}(V)$ is true usually, $j_\rho(\bm{x}):=\tilde{w}_\rho(\bm{x})/V$ takes a finite value for large $V$. 
Then, by leaving the terms of $\mathcal{O}(1/V)$ in Eq.~\eqref{sse}, we find the following Fokker--Planck equation: 
\begin{align}
    \pdv{\tilde{p}}{t}=-\sum_{\rho\in\mathscr{R}\cup\bar{\mathscr{R}}}
    \sum_{i\in\mathscr{S}_\X} \pdv{x_i}[\st_{i\rho}j_\rho(\bm{x})\tilde{p}(t,\bm{x})]
    +\frac{1}{2V}\sum_{\rho\in\mathscr{R}\cup\bar{\mathscr{R}}}\sum_{i,k\in\mathscr{S}_\X}
    \pdv[2]{}{x_i}{x_k}[\st_{i\rho}\st_{k\rho}j_\rho(\bm{x})\tilde{p}(t,\bm{x})] \label{fp}
\end{align}

Next we consider the relation between $j_\rho(\bm{x})$ and the reaction rates $J_\rho$. If we further take the limit of $V\to\infty$ in Eq.~\eqref{fp}, the Fokker--Planck equation leads to the Liouville equation 
\begin{align}
    \pdv{\tilde{p}}{t}=-\sum_{\rho\in\mathscr{R}\cup\bar{\mathscr{R}}}
    \sum_{i\in\mathscr{S}_\X} \pdv{x_i}[\st_{i\rho}j_\rho(\bm{x})\tilde{p}(t,\bm{x})], \label{liouville}
\end{align}
which describes the deterministic process given by the following ordinary differential equation~\cite{gardiner2009stochastic}: 
\begin{align}
    \dv{\bm{x}(t)}{t}=\sum_{\rho\in\mathscr{R}\cup\bar{\mathscr{R}}}\st_\rho j_\rho(\bm{x}(t)).  \label{rateeq}
\end{align}
Because $\st_{\bar{\rho}}=-\st_{\rho}$ for $\rho\in\mathscr{R}$, this equation is rewritten as 
\begin{align}
    \dv{\bm{x}(t)}{t}=\sum_{\rho\in\mathscr{R}}\st_\rho (j_\rho-j_{\bar{\rho}}), \label{rateeq2}
\end{align}
which corresponds to the rate equation~\eqref{rateeq_maintext}. Namely, $j_\rho$ is identified as the forward reaction rate $J_\rho^+$ if $\rho\in\mathscr{R}$, or the backward one $J_{\rho-M}^-$ if $\rho\in\bar{\mathscr{R}}$. 

The second term in the right hand side of Eq.~\eqref{fp} can be also represented by the reaction rates as 
\begin{align}
    \frac{1}{2V}\sum_{\rho\in\mathscr{R}}\sum_{i,k\in\mathscr{S}_\X}
    \pdv[2]{}{x_i}{x_k}[\st_{i\rho}\st_{k\rho}(J_\rho^++J_\rho^-)\tilde{p}(t,\bm{x})], 
\end{align}
where the sign in front of $J_\rho^-$ is plus because $\st_{i\bar{\rho}}\st_{k\bar{\rho}}=\st_{i\rho}\st_{k\rho}$. 
Therefore, the diffusion coefficient of the Fokker--Planck equation is obtained as 
\begin{align}
    D_{ik}=\frac{1}{2V}\sum_{\rho\in\mathscr{R}}\st_{i\rho}\st_{k\rho}(J_\rho^++J_\rho^-). 
\end{align}

\section{Note on the relation between the TSL and the information geometric SL}
We have derived another SL, which we call an information geometric speed limit (IGSL), in the previous study~\cite{yoshimura2021information}: 
\begin{align}
    \tau\geq \frac{\L^2}{2\C}=:\tau_\mathrm{IG}, \label{IG}
\end{align}
where $\mathcal{L}:=\int_0^\tau \dd{t}\sqrt{\dv*{s^2}{t^2}}$ is the length of the path, $\mathcal{C}:=(1/2)\int_0^\tau \dd{t}\dv*{s^2}{t^2}$ is the thermodynamic cost, and $\dv*{s^2}{t^2}=\sum_{i\in \mathscr{S}_\X} [\X_i]^{-1}(\!\dv*{[\X_i]}{t}\!)^2$ is the generalized Fisher information~\cite{ito2018stochastic,ito2020stochastic,yoshimura2021information}.
In this section, we compare the IGSL with the TSL. 

The IGSL is not always a better bound than the TSL, and vice versa. However, if the dynamics include completely irreversible reactions, the TSL will be useless because the entropy production diverges. On the other hand, the IGSL is still meaningful because the Fisher information can be defined whether there is an irreversible reaction. 

\begin{figure}
    \centering
    \includegraphics[width=0.5\linewidth]{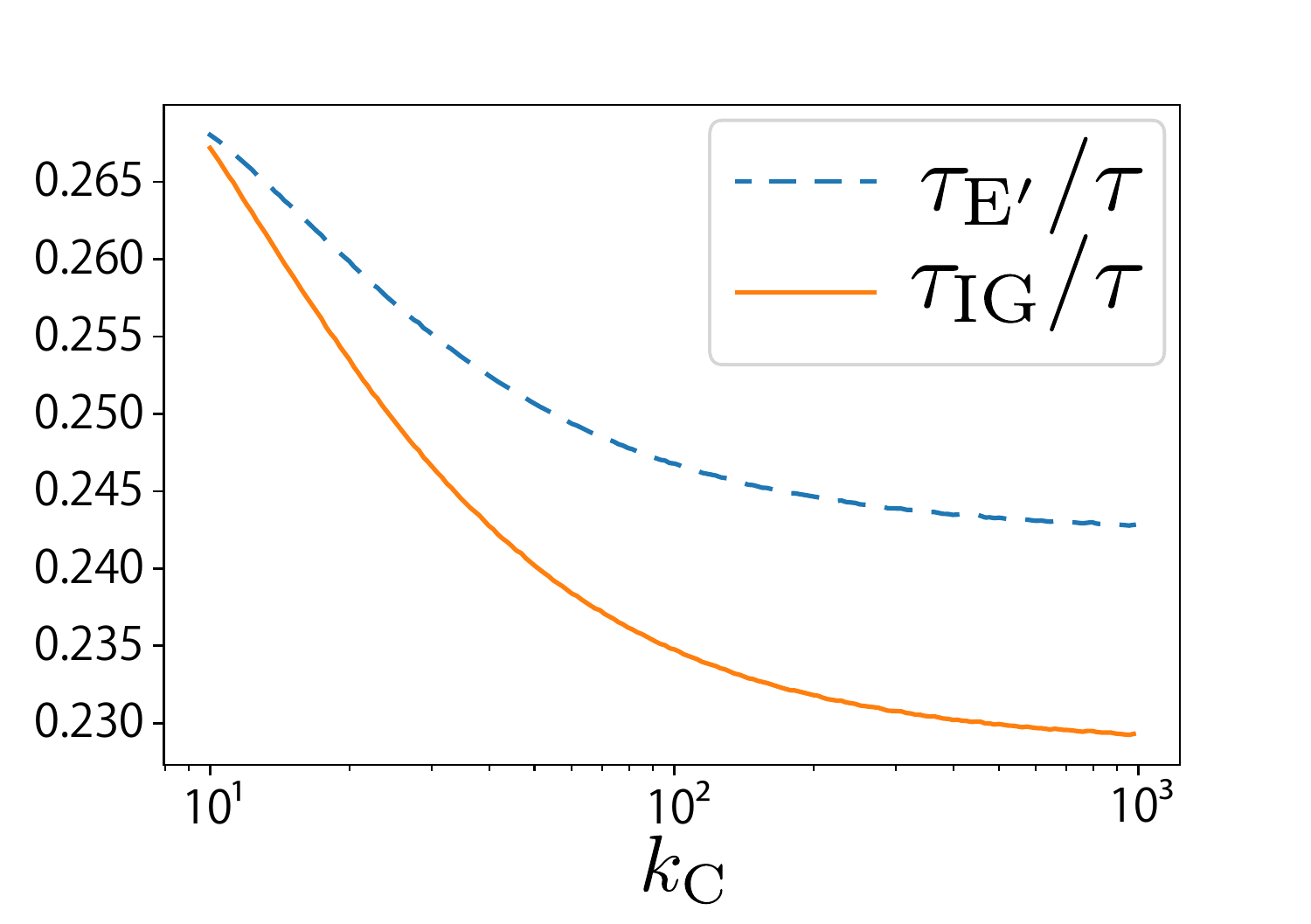}
    \caption{Comparison of the IGSL with the TSL. The CRN is the same as the one used in the second example in the main text. The IGSL is comparable to $\tau_\mathrm{E'}$, thus is much better than the other TSLs shown in the inset of Fig.~\ref{tsl_fig}. }
    \label{SM_IG_tsl}
\end{figure}

The less number of reactions is included in $\mathscr{R}_\S$, the tighter the TSL $\tau_\S$ will be because the main inequality in Eq.~\ref{mainineq} is proved by using the Cauchy--Schwarz inequality. 
On the other hand, the IGSL is tight when $\sqrt{\dv*{s^2}{t^2}}$ is constant, regardless of the number of reactions~\cite{yoshimura2021information}. 
In fact, the TSLs in Fig.~\ref{tsl_fig} other than $\tau_\mr{E'}$ are quite loose, which can be attributed to the fact that $\mathscr{R}_\S\;(\S\neq\{\mr{E'}\})$ contains multiple reactions. 
We compare $\tau_\mr{E'}/\tau$ and $\tau_\mr{IG}/\tau$ in Fig.~\ref{SM_IG_tsl} under the same setup as the second example in the main text. 
Although $\tau_\mr{IG}$ contains all of the contributions of $\S^\X$, it is as good a bound as $\tau_\mr{E'}$. 
If there are no species that are involved in only one reaction like $\mr{E'}$, the IGSL is always expected to work better than the TSL. 

It is also notable that the IGSL can be calculated as long as a time series of concentrations is available, so it is experimentally easier to obtain than the TSL, which requires a calculation of the entropy production. 

So far, we have introduced the aspects in which the IGSL is important. In the following, we will discuss the importance of the TSL in comparison with the IGSL. 
The IGSL uses the action function $\mathcal{C}$, which is the integral of the Fisher information, as a thermodynamic cost, but the physical meaning of the action function is not always clear. 
Under near-equilibrium conditions, 
\begin{align}
    \mathcal{C}=\frac{\dot{\Sigma}_{t=0}-\dot{\Sigma}_{t=\tau}}{4}
\end{align}
holds~\cite{yoshimura2021information}, so the action function is connected with a thermodynamic quantity. However, except in such special cases, we have not reached a general understanding of $\mathcal{C}$ that goes beyond the one as an abstract ``action function''. 
Conversely, the cost in the TSL is the fluctuation and the entropy production, which are quite intuitive costs. Therefore, the physical implication of the TSL is more obvious than that of the IGSL. 

\section{Details of numerical simulations}
We describe the models and parameters used in the examples in the main text in detail. 

When we simulate the Lotka--Volterra model~\eqref{LotkaVolterra}, we numerically solved the rate equation
\begin{align}
    \dv{[\mathrm{X}]_t}{t}&=J_1-J_2,\\
    \dv{[\mathrm{Y}]_t}{t}&=J_2-J_3,\\
    \dv{[\mathrm{A}]_t}{t}&=-J_1,\\
    \dv{[\mathrm{B}]_t}{t}&=J_2,
\end{align}
with the reaction rates
\begin{align}
    J_1&=k_1^+[\X]_t[\A]_t-k_1^-[\X]_t^2,\\
    J_2&=k_2^+[\X]_t[\Y]_t-k_2^-[\Y]_t^2,\\
    J_3&=k_3^+[\Y]_t-k_3^-[\mathrm{B}]_t,
\end{align}
where we set $k_1^+=10^{-4},k_2^+=2\times 10^{-1},k_3^+=10^{-1},$ and $k_1^-=k_2^-=k_3^-=10^{-3}$ and the initial conditions are $[\X]_0=1,[\Y]_0=10^{-1},[\A]_0=10^2,$ and $[\mathrm{B}]_0=10^{-3}$. 

For the enzymatic reaction with coenzyme~\eqref{enzyme}, we numerically solved the rate equation
\begin{align}
    \dv{[\mathrm{E}]_t}{t}&=-J_1+J_2-J_3,\\
    \dv{[\mathrm{ES}]_t}{t}&=J_1-J_2,\\
    \dv{[\mathrm{E'}]_t}{t}&=J_3,\\
    \dv{[\mathrm{C}]_t}{t}&=J_3+\mathcal{J}_\mathrm{C}^\Y,
\end{align}
with the reaction rates
\begin{align}
    J_1&=k_1^+[\mr{S}][\mr{E}]_t-k_1^-[\mr{ES}]_t\\
    J_2&=k_2^+[\mr{ES}]_t-k_2^-[\mr{P}][\mr{E}]_t\\
    J_3&=k_3^+[\mr{E}]_t-k_3^-[\mr{E}']_t[\mr{C}]_t
\end{align}
where $k_1^+=k_3^+=k_3^-=10,k_1^-=5\times 10^2,k_2^+=10^{-1}$ and $k_2^-=10^{-3}$. 
The initial concentrations of chemostatted species are $[\mr{S}]=1\times 10^3$, $[\mr{P}]=1$ and $[\mr{C}]=1\times 10^{-1}$, and the external concentration of $\mr{C}$ is set to $[\mathrm{C}]^\mr{ext}=2.1\times10^{-1}$.  Those of internal species are given by the steady-state distribution under the constraint $[\mr{E}]+[\mr{ES}]+[\mr{E'}]=1$. 
With a sufficiently large time $T=10$, 
we define $\tau$ as the time such that for all $t$ after it, 
\begin{align}
    \sum_{\X\in\{\mathrm{E},\mathrm{ES},\mathrm{E'}\}}\frac{|[\X]_t-[\X]_T|}{[\X]_T}<10^{-4}
\end{align}
is satisfied. 

\end{document}